\documentclass{article}
\usepackage{arxiv}
\usepackage[utf8]{inputenc} 
\usepackage[T1]{fontenc}    
\usepackage{hyperref}       
\usepackage{url}            
\usepackage{booktabs}       
\usepackage{amsfonts}       
\usepackage{nicefrac}       
\usepackage{microtype}      
\usepackage{lipsum}

\title{Trust Assessment in Online Social Networks\thanks{This paper is the full version of~\cite{8714017}, which is published in \textit{IEEE Transaction on Dependable and Secure Computing}.}}

 \author{
  Guangchi~Liu \\
  Research \& Development Department\\
  Stratifyd, Inc.\\
  Charlotte, NC, 28202 \\
  \texttt{luke.liu@stratifyd.com} \\
   \And
 Qing~Yang \\
  Department of Computer Science and Engineering\\
  University of North Texas\\
  Denton, TX 76207 \\
  \texttt{qing.yang@unt.edu} \\
  \And
Honggang~Wang \\
Department of Electrical and Computer Engineering \\
  University of Massachusetts Dartmouth\\
  North Dartmouth, MA, 02747 \\
  \texttt{hwang1@umassd.edu} \\
  \And
   Alex~X.~Liu \\
   Department of Computer Science and Engineering \\
   Michigan State university\\
   East Lansing, MI, USA \\
   \texttt{alexliu@cse.msu.edu} \\
}

\date{}
\usepackage{nccfoots}
\usepackage{balance}
\usepackage[usenames,dvipsnames]{color}
\usepackage[tight,TABTOPCAP]{subfigure}
\usepackage{caption2}
\usepackage[cmex10]{amsmath}
\usepackage{multirow}
\usepackage{algorithmic}
\usepackage{color, soul}
\usepackage{amsmath}
\usepackage{graphicx}
\usepackage{epstopdf}
\usepackage{soul}
\usepackage{comment}
\mathchardef\Gamma="0100 \mathchardef\Delta="0101
\mathchardef\Theta="0102 \mathchardef\Lambda="0103
\mathchardef\Xi="0104 \mathchardef\Pi="0105
\mathchardef\Sigma="0106 \mathchardef\Upsilon="0107
\mathchardef\Phi="0108 \mathchardef\Psi="0109
\mathchardef\Omega="010A

\newcommand{\outline}[1]{}

\usepackage{xspace}
\usepackage{url}
\usepackage{graphicx}
\usepackage{latexsym}
\usepackage{psfrag}
\usepackage[tight]{subfigure}
\usepackage{wrapfig}
\usepackage{comment}
\usepackage[ruled,linesnumbered,vlined]{algorithm2e}
\usepackage{alltt}
\usepackage{color}
\usepackage{setspace}
\usepackage{rotating}
\usepackage{enumerate}

\newcommand{\ie}{\emph{i.e.}\xspace}
\newcommand{\eg}{\emph{e.g.}\xspace}

\newtheorem{corollary}{Corollary}[section]
\newtheorem{theorem}{Theorem}[section]

\newcommand{\Comment}[1]{}

\usepackage{amsfonts}
\usepackage{graphicx}
\numberwithin{equation}{section}

\newtheorem{definition}{Definition}
\newtheorem{proof}{Proof}

\usepackage{xcolor}

\begin{document}





\maketitle


\begin{abstract}
Assessing trust in online social networks (OSNs) is critical for many applications such as online marketing and network security. It is a challenging problem, however, due to the difficulties of handling complex social network topologies and conducting accurate assessment in these topologies. To address these challenges, we model trust by proposing the three-valued subjective logic (3VSL) model. 3VSL properly models the uncertainties that exist in trust, thus is able to compute trust in arbitrary graphs. We theoretically prove the capability of 3VSL based on the Dirichlet-Categorical (DC) distribution and its correctness in arbitrary OSN topologies. Based on the 3VSL model, we further design the AssessTrust (AT) algorithm to accurately compute the trust between any two users connected in an OSN. We validate 3VSL against two real-world OSN datasets: Advogato and Pretty Good Privacy (PGP). Experimental results indicate that 3VSL can accurately model the trust between any pair of indirectly connected users in the Advogato and PGP. 
\end{abstract}




\section{Introduction}

%
Online social networks (OSNs) are among the most frequently visited places on the Internet. 
OSNs help people not only to strengthen their social connections with known friends but also to expand their social circles to friends of friends who they may not know previously.
Trust is the enabling factor behind user interactions in OSNs and is crucial to almost all OSN applications.
For example, in recommendation and crowdsourcing systems, trust helps to identify trustworthy opinions and/or users~\cite{Basu:2014:OPF:2670967.2670968}. 
%
%
In online marketing applications\cite{resnick2000reputation}, trust is used to identify trustworthy sellers. 
In a proactive friendship construction system \cite{yang2013maximizing}, trust enables the discovery of potential friendships. 
In wireless network domain, trust can help a cellular device to discover trustworthy peers to relay its data~\cite{wu2017social,8613914}.
In security domain, trust is considered an important metric to detect malicious users or websites~\cite{8057106, 6848107, 8438939}.
%
%
Given the above-mentioned applications, one confounding issue is to what degree a user can trust another user in an OSN.
This paper concerns the fundamental issue of trust assessment in OSNs: \emph{given an OSN, how to model and compute trust among users}?
Trust is traditionally considered as reputation or the probability of a user being benign. 
In online marketing, users rate each other based on their interactions, so the trust of a user can be derived from aggregated ratings. 
In the network security domain, however, the trust of a given user is defined as the probability that this user will behave normally in the future. 
Based on results from previous studies~\cite{rousseau1998not, mcknight2002developing}, we define trust as \emph{the probability that a trustee will behave as expected, from the perspective of a trustor}.
Here, both trustor and trustee are regular users in an OSN where the trustor is interested in knowing how trustworthy the trustee is. 
This general definition of trust makes it applicable for a wide range of applications.
We also assume that trust in OSNs is determined by objective evidence, i.e., cognition based trust~\cite{falcone2001social}, is not considered in this paper. 
\vspace{-0.15in}
\subsection{Problem Statements} 
We model a social network as a directed graph $G=(V,E)$ where a vertex $u \in V$ represents a user, and an edge $e(u,v) \in E$ denotes a trust relation from $u$ to $v$.
The weight of $e(u, v)$ denotes how much $u$ trusts $v$, which is commonly referred to as \textit{direct trust}.
A trustor may leverage the recommendations from other users to derive a trustee's trust, which is called \textit{indirect trust}.
We are interested in computing the indirect trust between two users who have not established a direct trust previously.
%
To solve this problem, we first need to design a trust model that works with both direct and indirect trust.
Based on the assumption that trust is determined by objective evidence, designing a trust model can be stated as follows.\\
\begin{itemize}
\vspace{-0.2in}
\item \textbf{P1}: \textit{Given the interactions between a trustor and a trustee, how to model the trust of the trustee, from the trustor's perspective?}
\end{itemize}
The second problem is to compute/infer indirect trust between users in an OSN.
Solving this problem means the trust between two users, without previous interactions, can be computed.
Because the indirect trust inference is available, a trustor can conduct a trust assessment of a trustee in an OSN.
As such, the second problem is formulated as follows.\\
\begin{itemize}
\vspace{-0.2in}
\item \textbf{P2}: \textit{Given a social network $G = (V,E)$, $\forall$ $u$ and $v$, s.t. $e(u,v)$ $\not\in E$ and $\exists$ at least one path from $u$ to $v$, how does one compute $u$'s trust in $v$, \ie, how should $u$ trust a stranger $v$?} 
\end{itemize}

\vspace{-0.15in}
\subsection{Proposed Approach}
\label{Proposed Approach}
To address problem \textbf{P1}, we propose the three-valued subjective logic (3VSL) model that accurately models the trust between a trustor and a trustee, based on their interactions.
3VSL is inspired by the subjective logic (SL) model~\cite{josang2001logic}, however, it is significantly different from SL.
%
%
%

The major difference between SL and 3VSL lies in the definitions of uncertainty in trust.
SL believes the uncertainty in the trust of a trustee never changes, however, 3VSL considers the uncertainty increases as trust propagates among users in an OSN.
Therefore, an extra state, called uncertainty state, is introduced in 3VSL to cope with the changing of uncertainty in trust.

The trust of a trustee, i.e., the probability that it will behave as expected, can be represented by a \textit{Dirichlet-Categorical} (DC) distribution that is characterized by three parameters $\alpha$, $\beta$ and $\gamma$.
Here, $\alpha$ is the number of positive interactions occurred, i.e., a trustor observed that the trustor behaved as expected for $\alpha$ times.
$\beta$ denotes the amount of negative interactions, indicating the trustee did not behave as expected.
It is also quite possible that the behavior of the trust is ambiguous, i.e., it is impossible to determine whether it behaved as expected or not.
In this case, we consider uncertain observations are made and use $\gamma$ to record them.
Uncertainty is generated not only when ambiguous behaviors are observed but also when trust propagates within an OSN, which will be elaborated in details in Section~\ref{CH:3vsl}.
The observations kept in $\alpha$, $\beta$ and $\gamma$ are also called \textit{evidence}, as they are used to judge whether the trustee is trustworthy or not.
The major reason of introducing the uncertain state in 3VSL is to accurately capture the trust propagation process.
When trust propagates from a user to another, certain evidence in $\alpha$ and $\beta$ are ``distorted''  and ``converted'' into uncertain evidence.
Given a DC distribution, it can be represented by a vector $\left< \alpha, \beta, \gamma \right>$, which is also called \textit{opinion}.
On the other hand, the trustee's trust can be derived from a DC distribution; therefore, trust can be represented by an opinion.
In the rest of this paper, we treat trust and opinion as interchangeable concepts, unless otherwise specified.

%
To address problem \textbf{P2}, we propose a trust assessment algorithm, called AssessTrust (AT), based on the 3VSL model.
%
%
The AT algorithm decomposes the network between the trustor and trustee as a parsing tree that provides the correct order of applying trust operations to computer the indirect trust between the two users.
Here, the trust operations available in trust computation are the discounting operation and combining operation.
Leveraging these two operations, AT is proven to be able to accurately compute the trust between any two users connected in an OSN. 
%
%
Because 3VSL appropriately treats the uncertainty in trust, AT offers more accurate trust assessments, compared to the topology- and graph-based solutions.
On the other hand, as AT aims at computing indirect trust between users, it outperforms the probability based models that focus only on direct trust.
Experiment results demonstrate that AT achieves the most accurate trust assessment results.
Specifically, AT achieves the F1 scores of $0.7$ and $0.75$, using the Advogato and Pretty Good Privacy (PGP) datasets, respectively.
AT can rank users based on their trust values.
We measure the accuracy of the ranking results, using the Kendall's tau coefficients.
Experiment results show that, on average, AT offers $0.73$ and $0.77$ kendall's tau coefficients, in Advogato and PGP, respectively.
\vspace{-0.15in}
\subsection{Technical Challenges and Solutions}
The first technical challenge is that 3VSL needs to accurately model the trust propagation and fusion in OSNs.
This is a challenge because trust propagation in OSNs is not well understood, although it is widely adopted by the research community.  
We address this challenge by using an opinion to represent trust and modeling trust propagation based on DC distribution and several commonly-accepted assumptions.
%


The second technical challenge is that 3VSL must be able to work on OSNs with non-series-parallel network topologies.
This is a challenge because the only allowed operations in trust assessment are trust propagation and trust fusion. 
However, these two operations require that a network's topology must be either series and/or parallel. 
This requirement cannot be satisfied in real-world online social networks.
We address this challenge by differentiating distorting opinions from original opinions. 
For example, if Alice trusts Bob and Bob trusts Charlie, then Alice's opinion on Bob is called the distorting opinion, and Bob's opinion on Charlie is the original opinion. 
We find that original opinions can be fused only once but distorting opinions can be combined any number of times. 
This discovery lays the foundation for the proposed recursive AssessTrust algorithm.
The third technical challenge is that 3VSL needs to handle social networks with arbitrary topologies, even with cycles.
This is a challenge because it is impossible to test 3VSL in all possible network topologies.
We address this challenge by mathematically proving 3VSL works in arbitrary networks. 
The proof is based upon the characteristics of Dirichlet distribution and the properties of different opinions in the trust computation process.
In the end, the AssessTrust algorithm is designed to compute the trust between any two users in an OSN.

The rest of this paper is organized as follows.
In Section~\ref{sec:bkgd}, the background and terminologies of trust are introduced.
In Section \ref{CH:3vsl}, we introduce the 3VSL model and define the trust propagation and fusion operations. 
We then differentiate discounting opinions from original opinions in Section \ref{CH:algorithm}, and prove 3VSL can handle arbitrary network topologies. 
In the same section, we detail the proposed AssessTrust algorithm.
In Section \ref{CH:eval}, we validate the 3VSL model and the AssessTrust algorithm, using two real-world datasets.
The related work is given in Section \ref{CH:RW}.
We conclude the paper in Section \ref{conclusion}.

\section{Background}
\label{sec:bkgd}

\subsection{Terminology}

In this section, we briefly introduce some terminologies frequently referred in this paper. 
Trust assessment is defined as the process that a \textit{trustor} assesses a \textit{trustee} on whether it will \textit{perform a certain task as expected}.
As such, trust can be either \textit{direct} or \textit{indirect}~\cite{7346861}.
Direct trust is formed from a trustor's direct interactions with a trustee while indirect trust is inferred from others' recommendations. 
%
Typically, trust is represented as an \textit{opinion}, indicating how much a trustor trusts a trustee.

To model trust propagation and trust fusion, two opinion operations, \ie, the discounting operation and combining operation, are design to facilitate trust computation/assessment~\cite{7346861}.
Trust fusion refers to combining different trust opinions to form a consensus trust opinion.
%
Trust propagation refers to a trust opinion being transferred from a user to another.
For example, if $A$ trusts $B$, and $B$ trusts $C$, then $B$'s opinion on $C$ will be discounted by $A$ to derive an indirect opinion of $C$'s trust.
%

\subsection{Subjective Logic}
To better understand 3VSL, we first briefly introduce the subjective logic (SL)~\cite{josang2001logic}. 
Considering two users $A$ and $X$, $A$'s opinion on $X$'s trust can be described by an \textit{opinion}.
\begin{equation*}
\begin{array}{l}
\omega _{AX}^{} = \left\langle {{\alpha _{AX}},{\beta _{AX}},2} \right\rangle \left| {{a_{AX}}}, \right.
\end{array}
\end{equation*}
where $\alpha _{AX}$, $\beta _{AX}$, and $2$ denote the amounts of evidence supporting user $X$ is trustworthy, untrustworthy, and uncertain, respectively. 
Based on $\alpha_{AX}$ and $\beta_{AX}$, a Beta distribution can be formed to model $A$'s trust in $X$.

In SL, the amount of uncertain evidence in an opinion is always $2$.
$a_{AX}$ is called the base rate and formed from existing impression without solid evidence, \eg, prejudice, preference, or a general opinion obtained from hearsay. 
For example, if $A$ always distrusts/trusts the users from a certain group where $X$ belongs to, then $a_{AX}$ will be smaller/greater than $0.5$. 
Note that the opinion in SL~\cite{josang2001logic} was defined as $\omega _{AX}^{} = \left\langle {{b _{AX}},{d_{AX}}, u_{AX}} \right\rangle \left| {{a_{AX}}} \right.$, which is another form of the trust opinion.
The connection between these two opinion representations is mentioned in~\cite{josang2001logic} as follows.
\begin{eqnarray}
\label{e4}
b_{AX} = \frac{\alpha_{AX}}{\alpha_{AX} + \beta_{AX} + 2}\nonumber\\
d_{AX} = \frac{\beta_{AX}}{\alpha_{AX} + \beta_{AX} + 2}\\
u_{AX} = \frac{2}{\alpha_{AX} + \beta_{AX} + 2}\nonumber
\end{eqnarray}

Leveraging the property of Beta distribution, two opinions can be fused, by combining the corresponding Beta distributions, to yield a new opinion.
For example, opinions ${\omega _1} = \left\langle {{\alpha _1},{\beta _1},2} \right\rangle \left| {{a_1}} \right.$ and ${\omega _2} = \left\langle {{\alpha _2},{\beta _2},2} \right\rangle \left| {{a_2}} \right.$ can be combined to produce ${\omega _{12}} = \left\langle {{\alpha _{12}},{\beta _{12}},2} \right\rangle \left| {{a_{12}}} \right.$, where $\alpha_{12}$, $\beta_{12}$ and $a_{12}$ are calculated as follows.
\begin{eqnarray*}
\left \{ \begin{array}{l}
{{\alpha _{12}} = {\alpha _1} + {\alpha _2}}\nonumber\\
{{\beta _{12}} = {\beta _1} + {\beta _2}}\nonumber\\
{{a_{12}} =  \displaystyle\frac{{{a_1} + {a_2}}}{2}}
\end{array}. \right.
\label{SLCOMB}
\end{eqnarray*}

Let $A$ and $B$ denote two users where ${\omega _1} = \left\langle {{\alpha _1},{\beta _1},2} \right\rangle \left| {{a_1}} \right.$ is $A$'s opinion about $B$'s trust.
Assume $C$ is another user and ${\omega_2} = \left\langle {{\alpha _2},{\beta _2},2} \right\rangle \left| {{a_2}} \right.$ is $B$'s opinion about $C$'s trust. 
Then, the discounting operation is applied to compute $A$'s indirect opinion about $C$'s trust $\omega_{AC} = \left\langle {{\alpha _{12}},{\beta _{12}},2} \right\rangle \left| {{a_{12}}} \right.$ where
\begin{eqnarray*}
 \left\{ \begin{array}{l}
{\alpha _{12}} = \displaystyle\frac{{{\alpha _1}{\alpha _2}}}{{\left( {{\beta _2} + {\alpha _2} + 2} \right)\left( {{\beta _1} + {\alpha _1} + 2} \right)}} \cdot \frac{2}{\kappa }\nonumber\\
{\alpha _{12}} = \displaystyle\frac{{{\alpha _1}{\beta _2}}}{{\left( {{\beta _2} + {\alpha _2} + 2} \right)\left( {{\beta _1} + {\alpha _1} + 2} \right)}} \cdot \frac{2}{\kappa }\nonumber\\
{a_{12}} = {a_2}
\end{array}, \right.
\label{SLDISC}
\end{eqnarray*}
and
\[\kappa  = 1 - \frac{{\left( {{\alpha _1}{\alpha _2} + {\alpha _1}{\beta _2}} \right)}}{{\left( {{\beta _2} + {\alpha _2} + 2} \right)\left( {{\beta _1} + {\alpha _1} + 2} \right)}}.\]

\section{Three-Valued Subjective Logic} 
\label{CH:3vsl}

The major limitation of the SL model is that the uncertainty in trust is considered a constant, however, the uncertainty in a trust opinion will be increased when it propagates from a user to another.
To address this issue, we propose the three-valued subjective logic (3VSL) to model trust between users in an OSN, by redefining the uncertainty in trust.
Designing the 3VSL model is a challenging task as trust propagation in OSNs is not well understood, although it is widely used in many applications.  
We address this challenge by modeling trust as an opinion, a representation of a probabilistic distribution over three different states, \ie, trustworthy, untrustworthy, and uncertain.
By investigating how these states of an opinion change during trust propagation, we redesign the trust discounting operation.
Leveraging the Dirichlet distribution, we also redesign the combining operation.
Moreover, we discover the mechanism of how to correctly apply these opinion operations on trust assessment within an OSN, leading to the design of the AssessTrust algorithm.

\subsection{A Probabilistic Interpretation of Trust}

Trust in 3VSL is defined as the probability that a trustee will behave as expected in the future. 
The probability is determined by the amounts of evidence that a trustor observed about a trustee's historical behaviors. 
A trustee may be observed behaving as expected, not expected, or in an ambiguous way. 
As a result, a trustor obtains positive, negative, and uncertain evidence accordingly.
Based on the observed evidence, Bayesian inference is used to infer the probability of a trustee being trustworthy, or the probability that a trustee will behave as expected in the future. 
In summary, given more positive observed evidence, the probability of a trustee being trustworthy is larger.

The uncertainty state in 3VSL not only contains the observed uncertain evidence but also the distorted evidence when trust propagates in the network. 
Knowing how much evidence is distorted will give us an idea of how much positive (and negative) evidence left, which must be accurate so that the probability inference (of trust) could be precise. 
Without keeping track of uncertainty evidence, the amount of certain evidence in an opinion becomes incorrect, leading to erroneous trust assessments. 
A trustee's future behavior can be modeled as a random variable $x$ that takes on one of three possible outcomes $\{1, 2, 3\}$, \ie, $x=1$, $x=2$ and $x=3$ indicating the trustee will behave as expected, not as expected, or in an ambiguous way, respectively.
As such, we are interested in the probability that $x=1$, which is determined by the positive observed behaviors of the trustee.
Therefore, the probability density function (pdf) of $x$ follows the Categorical distribution.
\begin{equation}
\label{eq:cat}
    f(x | \mathbf{p}) = \prod_{i=1}^3 p_i^{[x=i]}\nonumber,
\end{equation}
where $\mathbf{p} = (p_1, p_2, p_3)$ and $p_1 + p_2 + p_3 = 1$, $p_i$ represents the probability of observing event $i$. 
The Iverson bracket $[x=i]$ evaluates to 1 if $x = i$, and 0 otherwise.

If the value of $\mathbf{p}$ is available, the pdf of $x$ will be known and the probability of $x=i$ can be computed.
Unfortunately, $\mathbf{p}$ is an unknown parameter and needs to be estimated based on the observations of $x$.
We treat $\mathbf{p}$ as three random variables that follow the Dirichlet distribution.
\begin{equation}
\mathbf{p} \sim Dir(\alpha, \beta, \gamma)\nonumber,
\end{equation}
where $\alpha, \beta, \gamma$ are hyper-parameters that control the shape of the Dirichlet distribution. 
We assume $\mathbf{p}$ follows Dirichlet distribution mainly because it is a conjugate prior of categorical distribution.
In addition, because Dirichlet distribution belongs to a family of continuous multivariate probability distributions, we have various pdfs for $\mathbf{p}$ by changing the values of $\alpha, \beta, \gamma$.
\begin{equation}
\label{eq:fp}
f(\mathbf{p}) = C {p_1}^{\alpha-1} {p_2}^{\beta-1} {p_3}^{\gamma-1},
\end{equation}
where $C$ is a normalizing factor ensuring $p_1 + p_2 + p_3 = 1$. 
In this way, we use $\mathbf{p} \sim Dir(\alpha, \beta, \gamma)$ to model the uncertainty in estimating $\mathbf{p}$.

With the mathematical model in place, $\mathbf{p}$ can be estimated based on the observations of $x$, according to the Bayesian inference.
Given a set of independent observations of $x$, denoted by $\mathbf{D} = \{x_1, x_2, \cdots, x_n\}$ where $x_j \in \{1, 2, 3\}$ and $j=1, 2, \cdots, n$, we want to know how likely $\mathbf{D}$ is observed.
This probability can be computed as
\begin{equation}
P(\mathbf{D}|\mathbf{p}) = \prod_{j=1}^{n} p_1^{[x_j = 1]} p_2^{[x_j = 2]} p_3^{[x_j=3]}.\nonumber
\end{equation}
Let $c_i$ denote the number of observations where $x=i$, then the above equation becomes $p_1^{c_1} p_2^{c_2} p_3^{c_3}$. 
Based on Bayesian inference, given observed data $\mathbf{D}$, the  posterior pdf of $\mathbf{p}$ can be estimated from
\begin{equation}
f(\mathbf{p} | \mathbf{D}) = \frac{P(\mathbf{D} | \mathbf{p}) f(\mathbf{p})}{P(\mathbf{D})},\nonumber
\end{equation}
where $P(\mathbf{D} | \mathbf{p}) = p_1^{c_1} p_2^{c_2} p_3^{c_3}$ is the likelihood function, and $f(\mathbf{p})$ the prior pdf of $\mathbf{p}$.
$P(\mathbf{D})$ is the probability that $\mathbf{D}$ is observed, which is independent of $\mathbf{p}$. 
Therefore, we have
\begin{equation}
f(\mathbf{p} | \mathbf{D}) \propto p_1^{c_1} p_2^{c_2} p_3^{c_3} \times p_1^{\alpha-1} p_2^{\beta - 1} p_3 ^{\gamma -1}.\nonumber
\end{equation}
That means the posterior pdf $f(\mathbf{p} | \mathbf{D})$ can be modeled by another Dirichlet distribution $Dir(\alpha + c_1, \beta + c_2, \gamma + c_3)$.
With the posterior pdf of $\mathbf{p}$, we have the following predicative model for $x$.
\begin{equation}
\label{eq:dc}
f(x|\mathbf{D}) = \int f(x | \mathbf{p}) f(\mathbf{p} | \mathbf{D})d\mathbf{p}.
\end{equation}
This function is in fact a composition of Categorical ($f(x | \mathbf{p})$) and Dirichlet ($f(\mathbf{p} | \mathbf{D})$) distributions, so it is called Dirichlet-Categorical (DC) distribution~\cite{tu2014dirichlet}.

\subsection{Opinion}

In the previous section, we introduce how to model a trustee's future behavior by a DC distribution.
From a DC distribution, the probability that the trustee is trustworthy can be derived from Eq.~\ref{eq:dc}.
%
Because the shape of a DC distribution is determined by three parameters, we use these parameters to form a vector to represent it.
This vector is called \textit{opinion} that expresses a trustor's opinion about a trustee's trust.

For a given DC distribution, the only undetermined parameters are $\alpha, \beta, \gamma$.
We set $\alpha = \beta = \gamma =1$, if there is no observed data, \ie, $\mathbf{D} = \emptyset$.
In this case, the DC distribution yields a uniform distribution, \ie, $p_1 = p_2 = p_3 = 1/3$. 
Assuming $\mathbf{p}$ initially follows uniform distribution is reasonable because we make no observation of $x$, and the best choice is to believe that $x$ could be $1$, $2$, or $3$ with equal probability.
As more observations of $x$ are made, the pdf of $\mathbf{p}$ becomes more accurate.

From Eq.~\ref{eq:dc}, we can compute the probability of $x = 1$, \ie, whether a trustee will behave as expected.
In other words, we can use Eq.~\ref{eq:dc} to infer the trust of the trustee.
Specifically, we can obtain the expectation of the probability that the trustee will behave as expected as follows.
\begin{align}
      &P(x = 1 | \mathbf{D}) \nonumber\\
 	  &= \int P(x=1 | p_1, p_2, p_3) P (p_1, p_2, p_3 | c_1, c_2, c_3) d(p_1, p_2, p_3)  \nonumber\\
      &= \frac{\Gamma (c_1 + c_2 + c_3)}{\Gamma(c_1) \Gamma(c_2) \Gamma(c_3)} \int p_1^{c_1 - 1} p_2^{c_2 - 1} p_3^{c_3 - 1}  \nonumber\\
      & = \frac{\Gamma(c_1 + c_2 + c_3) \Gamma(c_1 +1) \Gamma(c_2) \Gamma(c_3)}{\Gamma(c_1) \Gamma(c_2) \Gamma(c_3) \Gamma(c_1 + c_2 + c_3+1)}  \nonumber\\
      & = \frac{c_1}{c_1 + c_2 + c_3},
      \label{expb_pos}
\end{align}
where $\Gamma(n) = (n-1)!$ is the Gamma function.
In the same way, the probabilities that the trustee will behave not as expected, or in an ambiguous way, can be computed from 
\[
P(x = 2 | \mathbf{D}) = \frac{c_2}{c_1 + c_2 + c_3},
\]
and 
\[
P(x = 3 | \mathbf{D}) = \frac{c_3}{c_1 + c_2 + c_3}.
\]

If the hyper-parameters $\alpha, \beta, \gamma$ equal to 1, the future behavior of the trustee is only determined by $c_1, c_2, c_3$, \ie, the numbers of observations collected when the trustee behaved as expected, not as expected, or in an ambiguous way. 
We name these observations as positive, negative, and uncertain evidence.
From a trustor $A$'s perspective, a trustee $X$'s future behavior can be modeled a DC distribution that is represented as an opinion.
\begin{equation}
\omega_{AX} = \left<\alpha_{AX}, \beta_{AX}, \gamma_{AX}\right> | a_{AX}.\nonumber
\end{equation}
Here, $\omega_{AX}$ denotes $A$'s opinion on $X$'s future behavior, or $A$'s trust in $X$ behaving as expected.
The parameters $\alpha_{AX}, \beta_{AX}, \gamma_{AX}$ refer to the amounts of observed positive, negative and uncertain evidence, respectively.
We further name them as the \textit{belief}, \textit{distrust} and \textit{uncertainty} parameters, in the rest of the paper.
The subscripts of $\alpha_{AX}, \beta_{AX}, \gamma_{AX}$ differentiate them from the prior $\alpha, \beta, \gamma$, \ie, the former represents observed evidence while the latter is always $(1, 1, 1)$.

\subsection{Discounting Operation}
\begin{figure}
\centering
\subfigure[A general illustration of series topology.]{\includegraphics[width=3in]{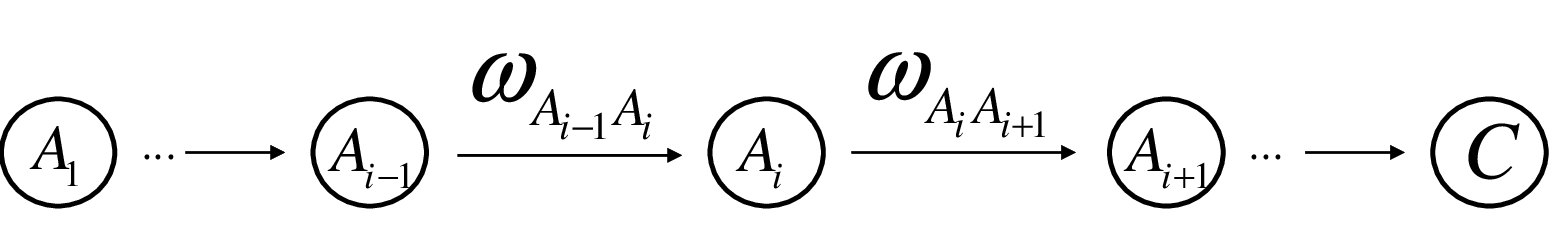}
\label{s-1}}
\subfigure[A simple example of series topology.]{\includegraphics[width=1.875in]{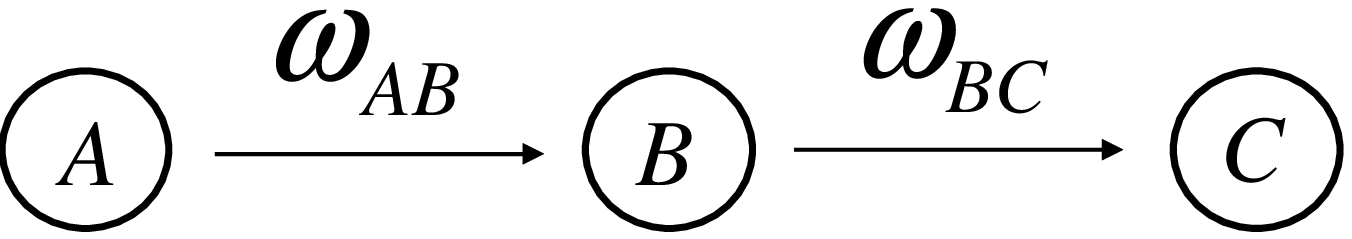}
\label{s-2}}
\caption{Examples of series topologies}
\label{series_topo}
\end{figure}

Trust propagation in OSNs was well-known, however, there is a lack of understanding about how to computationally model the process in practice.
Trust propagation can be illustrated by a series topology, as shown in Fig.~\ref{s-1}.
In the figure, two edges are connected in \textit{series} if they are incident to a vertex of degree $2$.
Trust propagation means that if user $A_{i-1}$ trusts $A_{i}$ and $A_{i}$ trusts $A_{i+1}$, then $A_{i-1}$ can derive an indirect trust of $A_{i+1}$, even if $A_{i-1}$ did not interact with $A_{i+1}$ before.

Based on existing literature on trust propagation~\cite{Guha:2004:PTD:988672.988727, Borgs2010, Ziegler2005, guha2004propagation}, it is commonly agreed that the following assumptions hold.
\begin{itemize}
\item A1: If $A$ trusts $B$, $B$ trusts $C$, then $A$ trusts $C$.
\item A2: If $A$ trusts $B$, $B$ does not trust $C$, then $A$ does not trust $C$.
\item A3: If $A$ trusts $B$, $B$ is uncertain about the trust of $C$, then $A$ is uncertain about $C$'s trust.
\item A4: If $A$ does not trust $B$, or $A$ is uncertain about $B$, then $A$ is uncertain about the trust of $C$.
\end{itemize} 
It is worth mentioning that if $A$ does not trust or is uncertain about $B$, then $A$ is uncertain about $C$, and $B$'s opinion on $C$ cannot propagate to $A$.
Based on the above-mentioned four assumptions, the trust propagation process can be modelled by the logic operation on two trust opinions.

Let's denote $A$'s opinion on $B$ as
\[
\omega_{AB} = \left<\alpha_{AB}, \beta_{AB}, \gamma_{AB}\right>,
\]
and $B$'s opinion on $C$ as
\[
\omega_{BC} = \left<\alpha_{BC}, \beta_{BC}, \gamma_{BC}\right>,
\]
where $\{\alpha_{AB}, \beta_{AB}, \gamma_{AB}\} = \mathbf{D}_{AB}$ and $\{\alpha_{BC}, \beta_{BC}, \gamma_{BC}\} = \mathbf{D}_{BC}$ represent the observations made by $A$ and $B$, about $B$ and $C$, respectively.
In this way, the expected probability that $C$ will behave as what $A$ expects will be 
\begin{eqnarray}
&&\iint (x = 1|{{\bf{p}}_{AB}})f({{\bf{p}}_{AB}}|{{\bf{D}}_{AB}}) \times \nonumber\\
&&f(x = 1|{{\bf{p}}_{BC}})f({{\bf{p}}_{BC}}|{{\bf{D}}_{BC}})d({{\bf{p}}_{AB}})d({{\bf{p}}_{BC}}).\nonumber\\
\label{Dis_Opr_b}  
\end{eqnarray}

The equation makes sense because $A$ trusts $C$ if and only if $A$ trusts $B$ and $B$ trusts $C$, which is the assumption A1.
In other words, the probability that $C$ will behave as what $A$ expects is equal to the probability that $C$ will behave as what $B$ expects, if $A$ trusts $B$. 
In the above equation, $f(x = 1 | \mathbf{p}_{AB}) f(\mathbf{p}_{AB}| \mathbf{D}_{AB})$ gives the probability that $A$ trusts $B$, and $f(x = 1|\mathbf{p}_{BC}) f(\mathbf{p}_{BC} | \mathbf{D}_{BC}) d(\mathbf{p}_{AB})$ is the probability that $B$ trusts $C$.

Because the probabilities that $A$ trusts $B$ and $B$ trusts $C$ are independent of each other, Eq.~\ref{Dis_Opr_b} can be rewritten as
\begin{eqnarray}
  \int  f(x = 1 | \mathbf{p}_{AB}) f(\mathbf{p}_{AB}| \mathbf{D}_{AB}) d(\mathbf{p}_{AB})\times \nonumber\\
  \int f(x = 1| \mathbf{p}_{BC}) f(\mathbf{p}_{BC} | \mathbf{D}_{BC})  d(\mathbf{p}_{BC}).
\label{Dis_Opr_b_0}  
\end{eqnarray} 
The two integrals in the equation are used to compute the expected probabilities that $A$ trusts $B$ and $B$ trusts $C$, respectively.
According to Eq.~\ref{expb_pos}, we know 
\begin{eqnarray*}
&&\int  f(x = 1 | \mathbf{p}_{AB}) f(\mathbf{p}_{AB}| \mathbf{D}_{AB}) d(\mathbf{p}_{AB})  \nonumber\\
&=&\frac{\alpha_{AB}}{\alpha_{AB}+ \beta_{AB}+ \gamma_{AB}},  \nonumber\\
\end{eqnarray*} 
and
\begin{eqnarray*}
&& \int f(x = 1| \mathbf{p}_{BC}) f(\mathbf{p}_{BC} | \mathbf{D}_{BC})  d(\mathbf{p}_{BC}) \nonumber\\
& =& \frac{\alpha_{BC}}{\alpha_{BC}+ \beta_{BC}+ \gamma_{BC}}.
\end{eqnarray*} 
Inserting these two values into Eq.~\ref{Dis_Opr_b_0}, we have the probability that $C$ will behave as what $A$ expects as
\begin{equation}
\frac{\alpha _{AB} \alpha_{BC}} {(\alpha _{AB} + \beta_{AB} + \gamma_{AB}) (\alpha_{BC} + \beta_{BC} + \gamma_{BC})}.
\label{Dis_Opr_b_2}
\end{equation} 
According to assumption A2, the probability that $C$ will not behave as what $A$ expects is
\begin{eqnarray}
 && \iint f(x = 1 | \mathbf{p}_{AB}) f(\mathbf{p}_{AB}| \mathbf{D}_{AB}) \times \nonumber\\
 && f(x = 2| \mathbf{p}_{BC}) f(\mathbf{p}_{BC} | \mathbf{D}_{BC}) d(\mathbf{p}_{AB}) d(\mathbf{p}_{BC}).\nonumber\\
\label{Dis_Opr_n}
\end{eqnarray}
This is because $A$ does not trust $C$, if and only if $A$ trusts $B$ but $B$ does not trust $C$. 
Because these probabilities are independent, we have the expected probability that $A$ does not trust $C$ as
\begin{equation}
\frac{\alpha _{AB} \beta_{BC}} {(\alpha _{AB} + \beta_{AB} + \gamma_{AB}) (\alpha_{BC} + \beta_{BC} + \gamma_{BC})}.
\label{Dis_Opr_n_2}
\end{equation}

Finally, based on assumptions A3 and A4, the expected probability that $C$ will behave in an ambiguous way is 
\begin{equation}
\begin{split}
& \iint    f(x = 1 | \mathbf{p}_{AB}) f(\mathbf{p}_{AB}| \mathbf{D}_{AB})\times \\\nonumber
& f(x = 3| \mathbf{p}_{BC}) f(\mathbf{p}_{BC} | \mathbf{D}_{BC}) + \\\nonumber
&  f(x = 2 | \mathbf{p}_{AB}) f(\mathbf{p}_{AB}| \mathbf{D}_{AB}) + f(x = 3 | \mathbf{p}_{AB}) f(\mathbf{p}_{AB}| \mathbf{D}_{AB}) \\\nonumber
& d(\mathbf{p}_{AB}) d(\mathbf{p}_{BC}) . \\
\end{split}
\label{Dis_Opr_u}
\end{equation}
The expected probability can be rewritten as
\begin{equation}
\displaystyle \frac{\alpha _{AB} \gamma _{BC} + (\beta _{AB} + \gamma _{AB})(\alpha _{BC} + \beta _{BC} + \gamma _{BC}) } {(\alpha _{AB} + \beta _{AB} + \gamma _{AB}) (\alpha _{BC} + \beta _{BC} + \gamma _{BC})}.
\label{Dis_Opr_u_2}
\end{equation}

The summation of Eqs.~\ref{Dis_Opr_b},~\ref{Dis_Opr_n} and~\ref{Dis_Opr_u} equals $1$.
The three equations actually give the estimated probabilities that $C$ will behave as expected, not as expected, or in an ambiguous way, respectively.
Putting these values back into the Categorical distribution
\begin{equation}
f(x | \mathbf{p}_{AC}) = \prod_{i=1}^3 p_i^{[x=i]},
 \label{eq:ac}
\end{equation}
where $\mathbf{p}_{AC} = (p_1, p_2, p_3)$ and
\begin{eqnarray}
&&{p_1} = \displaystyle \frac{{{\alpha _{AB}}{\alpha _{BC}}}}{{({\alpha _{AB}} + {\beta _{AB}} + {\gamma _{AB}})({\alpha _{BC}} + {\beta _{BC}} + {\gamma _{BC}})}}\nonumber\\
&&{p_2} = \displaystyle \frac{{{\alpha _{AB}}{\beta _{BC}}}}{{({\alpha _{AB}} + {\beta _{AB}} + {\gamma _{AB}})({\alpha _{BC}} + {\beta _{BC}} + {\gamma _{BC}})}},\nonumber\\
&&{p_3} = \displaystyle \frac{{({\beta _{AB}} + {\gamma _{AB}})({\alpha _{BC}} + {\beta _{BC}} + {\gamma _{BC}}) + {\alpha _{AB}}{\gamma _{BC}}}}{{({\alpha _{AB}} + {\beta _{AB}} + {\gamma _{AB}})({\alpha _{BC}} + {\beta _{BC}} + {\gamma _{BC}})}}\nonumber\\
\label{Dis_Opr_rst}
\end{eqnarray} 
we can compute the probability that $X$ is trustworthy.

From the above description, we know the Categorical distribution is in fact derived from $B$'s opinion on $C$.
Let's assume $B$ made the observations $\mathbf{x} = \{x_1, x_2, \cdots, x_n\}$ on $C$. 
Then, we know $\alpha_{BC}, \beta_{BC}, \gamma_{BC}$ equal to the numbers of observations where $x =1, x = 2, x = 3$, respectively.
Clearly, the observations $B$ made about $C$ do not reflect $A$'s opinion on $C$.
As trust propagates among users, $A$ could derive an indirect opinion from $B$'s observations of $C$.
One may ask the question that if $A$ was provided with the $n$ observations, how many of them will be considered positive, negative, and uncertain, from $A$'s perspective.
That implies$A$ needs to reuse $B$'s observations on $C$ to derive its own opinion on $C$.
For each $x_j \in \mathbf{x} $ where $j = 1, 2, \cdots, n$, we know $x_j$ is observed, given the underlying categorical distribution shown in Eq.~\ref{eq:ac}. 
In other words, $\mathbf{x}$ follows the multinomial distribution with parameters $(n, \mathbf{p}_{AC})$. 
From the multinomial distribution, we can recompute the probability that $C$ is trustworthy, by re-categorizing the observations $\mathbf{x}$.
We know the following re-categorization will occur with the highest probability.
\begin{eqnarray}
\alpha _{AC} &=& p_1 (\alpha _{BC} + \beta _{BC} + \gamma _{BC})  \nonumber\\
&=&\displaystyle \frac{\alpha _{AB} \alpha _{BC}}{(\alpha _{AB} + \beta _{AB} + \gamma _{AB})},\nonumber \\
\beta _{AC}&=& p_2 (\alpha _{BC} + \beta _{BC} + \gamma _{BC}) \nonumber \\
&=&\displaystyle \frac{\alpha _{AB} \beta _{BC}} {(\alpha _{AB} + \beta _{AB} + \gamma _{AB})},\nonumber\\
\gamma _{AC} &=& p_3 (\alpha _{BC} + \beta _{BC} + \gamma _{BC}),  \nonumber\\
&=&\displaystyle \frac{(\beta _{AB} + \gamma _{AB}) (\alpha _{BC} + \beta _{BC} + \gamma _{BC}) + \alpha _{AB}\gamma _{BC}} {(\alpha _{AB} + \beta _{AB} + \gamma _{AB})}.\nonumber\\
\label{Dist_Evid}
\end{eqnarray} 
Therefore, we use $\omega_{AC} = \left< \alpha_{AC}, \beta_{AC}, \gamma_{AC}\right>$ to represent $A$'s opinion about $C$'s trust.
Note that the opinion $\omega_{AC} $ is generated by distorting positive and negative evidence in $\omega_{BC}$ to uncertain evidence. 
That also means the total amount of evidence does not change in trust propagation.
\begin{eqnarray}
{\alpha _{AC}} + {\beta _{AC}} + {\gamma _{AC}} = {\alpha _{BC}} + {\beta _{BC}} + {\gamma _{BC}}.
\label{evid_unchange}
\end{eqnarray}
Based on the previous analysis, we formally define the discounting operation in 3VSL as follows.\\
\begin{definition}[Discounting Operation]
Given three users $A$, $B$ and $C$, if $\omega _{AB} = \left<\alpha_{AB}, \beta_{AB}, \gamma_{AB}\right>$ is $A$'s opinion on $B$'s trust, and $\omega _{BC} = \left<\alpha_{BC}, \beta_{BC}, \gamma_{BC}\right>$ is $B$'s opinion on $C$'s trust, the discounting operation $\Delta (\omega _{AB},\omega _{BC})$ computes $A$'s opinion on $C$ as
\[
\Delta ({\omega _{AB}},{\omega _{BC}}) = \left<{\alpha _{AC}},{\beta _{AC}},{\gamma _{AC}}\right>,
\]
where
\begin{eqnarray}
{\alpha _{AC}} &=& \displaystyle \frac{{{\alpha _{AB}}{\alpha _{BC}}}}{{({\alpha _{AB}} + {\beta _{AB}} + {\gamma _{AB}})}},\nonumber\\
{\beta _{AC}} &=& \displaystyle  \frac{{{\alpha _{AB}}{\beta _{BC}}}}{{({\alpha _{AB}} + {\beta _{AB}} + {\gamma _{AB}})}},\nonumber\\
{\gamma _{AC}} &=& \displaystyle  \frac{{({\beta _{AB}} + {\gamma _{AB}})({\alpha _{BC}} + {\beta _{BC}} + {\gamma _{BC}}) + {\alpha _{AB}}{\gamma _{BC}}}}{{({\alpha _{AB}} + {\beta _{AB}} + {\gamma _{AB}})}}.\nonumber\\
\label{dis_opr}
\end{eqnarray}
\end{definition}
Opinion $\omega_{BC}$ being discounted can be viewed as the certain evidence in $\omega_{BC}$ are distorted by opinion $\omega_{AB}$, and then transferred into the uncertainty space of $\omega_{AC}$. 
Because the total amount of evidence in opinion $\omega_{AC} = \Delta(\omega_{AB}, \omega_{BC})$ is the same as $\omega_{BC} $'s, we conclude \textit{the resulting opinion of discounting operation shares exactly the same evidence space as the original opinion}.

Based on the definition of discounting operation, it offers two interesting properties: decay and associative properties.
\begin{corollary}
Decay Property:
Given two opinions $\omega _{AB}$ and $\omega _{BC}$,  $\Delta(\omega_{AB}, \omega_{BC}) $ yields a new opinion $\omega_{AC}$, where $\alpha_{AC} \le \alpha_{BC}$, $\beta_{AC} \le \beta_{BC}$ and $\gamma_{AC} > \gamma_{BC}$.
\label{c4}
\end{corollary}

\begin{proof}
Because $\displaystyle\frac{{{\alpha _{AB}}}}{{({\alpha _{AB}} + {\beta _{AB}} + {\gamma _{AB}})}} \le 1$, according to Eq~\ref{dis_opr}, we have $\alpha_{AC} \le \alpha_{BC}$ as well as $\beta_{AC} \le \beta_{BC}$. Hence, $-\alpha_{AC} - \beta_{AC} \ge - \beta_{AC} - \beta_{AC}$. According to Eq.~\ref{evid_unchange}, we have $\gamma_{AC} \ge \gamma_{BC}$.
\end{proof}
In other words, by applying the discounting operation, the uncertainty in trust (or in the resulting opinion) increases. 
This property implies that the more trust propagates among users in an OSN, the more uncertain the resulting opinion.

\begin{corollary}
Associative Property:
Given three opinions $\omega _{AB}$, $\omega _{BC}$ and $\omega _{CD}$, $\Delta (\Delta (\omega _{AB},\omega _{BC}),\omega _{CD}) \equiv \Delta (\omega _{AB},\Delta (\omega _{BC},\omega _{CD}))$.
\label{c3}
\end{corollary}

\begin{proof}
Simply based on Eq~\ref{dis_opr}.
\end{proof}

However, the discounting operation is not commutative, \ie, $\Delta (\omega _{AB},\omega _{BC}) \neq \Delta (\omega _{BC},\omega _{AB})$. 
Given a series topology where opinions are ordered as $\omega _{A_1 A_2}, \omega _{A_2, A_3}, \cdots,$ $\omega _{A_{n-1} A_{n}}$, the final opinion can be calculated as $\Delta (\Delta (\Delta (\omega _{A_1 A_2},\omega _{A_2 A_3}), \cdots),\omega _{A_{n-1} A_{n}})$. 
As the discounting operation is associative, it can be simplified as $\Delta(\omega _{{A_1 A_2}},\omega _{A_2 A_3}, \cdots \omega _{A_{n-1} A_{n}})$.

\subsection{Combining Operation}
\begin{figure}
\centering
\subfigure[]{\includegraphics[width=1.5in]{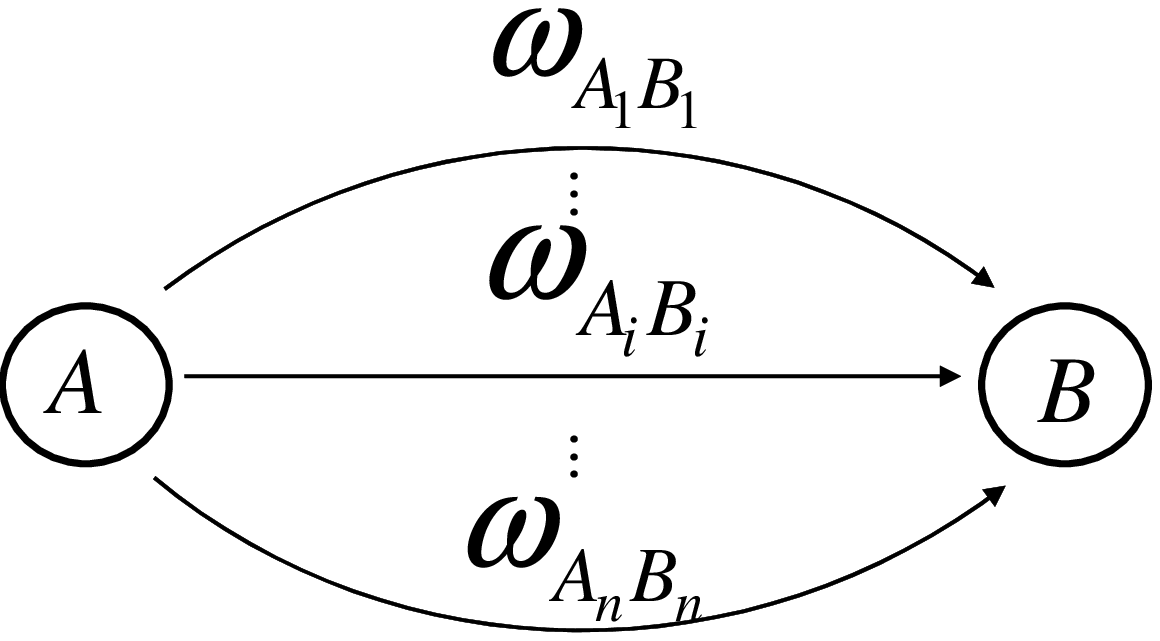}
\label{p-1}}
\subfigure[]{\includegraphics[width=1.5in]{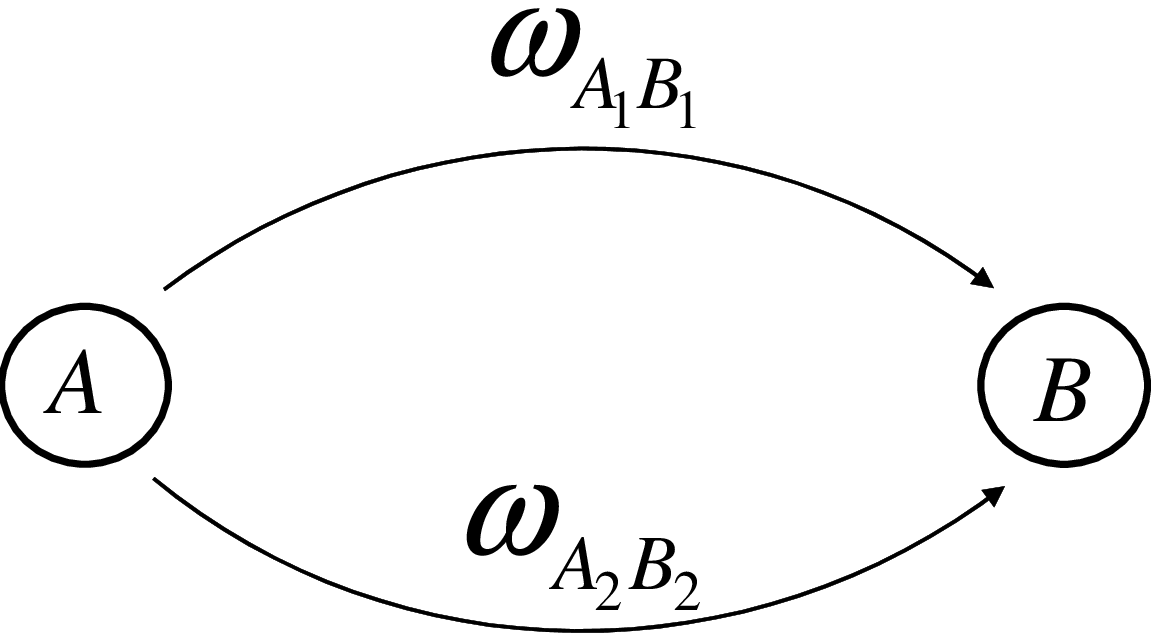}
\label{p-2}}
\caption{Examples of parallel topologies}
\label{parallel_topo}
\end{figure}

According to previous works~\cite{Guha:2004:PTD:988672.988727, Borgs2010, Ziegler2005}, trust opinions can be fused into a consensus one by aggregating the evidence from each opinion. 
We will use the parallel topology shown in Fig.~\ref{p-2} to explain how the combining operation works.

Let $\omega_{{A_1}{B_1}} = \left<\alpha_{{A_1}{B_1}}, \beta_{{A_1}{B_1}}, \gamma_{{A_1}{B_1}}\right>$ and $
\omega_{{A_2}{B_2}} = \left<\alpha_{{A_2}{B_2}}, \beta_{{A_2}{B_2}}, \gamma_{{A_2}{B_2}}\right>$ be $A$'s two indirect/direct opinions on $B$.
We use $ \{\alpha_{{A_1}{B_1}}, \beta_{{A_1}{B_1}}, \gamma_{{A_1}{B_1}}\} = \mathbf{D}_{{A_1}{B_1}}$ and $ \{\alpha_{{A_2}{B_2}}, \beta_{{A_2}{B_2}}, \gamma_{{A_2}{B_2}}\} = \mathbf{D}_{{A_2}{B_2}}$ to represent the two sets of observations $A$ made on $B$.
%
According to the definition of an opinion, the expected probability that $B$ will behave as what $A$ expects can be computed from the following DC distribution.
\begin{equation}
\int {f(x = 1|{{\bf{p}}_{AB}})f({{\bf{p}}_{AB}}|{{\bf{D}}_{{A_1}{B_1}}}, {{\bf{D}}_{{A_2}{B_2}}})d({{\bf{p}}_{AB}})},
\label{Com_Opr_b}  
\end{equation}
The intuition of Eq.~\ref{Com_Opr_b} can be explained as follows.
$A$ first infers the parameters ${\bf{p}}_{AB}$ by aggregating the observations ${\bf{D}}_{{A_1}{B_1}}$ and ${\bf{D}}_{{A_2}{B_2}}$, \ie, the posterior pdf of ${\bf{p}}_{AB}$ becomes
\begin{equation}
f({{\bf{p}}_{AB}}|{{\bf{D}}_{{A_1}{B_1}}}, {{\bf{D}}_{{A_2}{B_2}}}).
\label{comb_evid}
\end{equation}
Then, based on the inferred parameters ${\bf{p}}_{AB}$, the probability that $B$ will behave as what $A$ expects can be computed from $f(x = 1|{{\bf{p}}_{AB}})$.   
Considering all possible values of ${\bf{p}}_{AB}$, we can obtain the Eq.~\ref{Com_Opr_b}.

Now, we will give the analytic form of Eq.~\ref{comb_evid} as follows. 
$A$ first forms his opinion on $B$ from ${{\bf{D}}_{{A_1}{B_1}}}$.
As such, the pdf of ${{\bf{p}}_{{A}{B}}}$ is obtained.
Then, $A$ adjusts the estimate for ${{\bf{p}}_{{A}{B}}}$ based on another set of evidence ${{\bf{D}}_{{A_2}{B_2}}}$.
As a matter of fact, Eq.~\ref{comb_evid} can be regarded as the distribution of ${{\bf{p}}_{{A}{B}}}$ based on (1) the posterior evidence in ${{\bf{D}}_{{A_2}{B_2}}}$ and (2) the prior parameters ${{\bf{p}}_{{A_1}{B_1}}}$ estimated from ${{\bf{D}}_{{A_1}{B_1}}}$.
According to Bayesian inference, it can be expressed as follows.
\begin{eqnarray}
&&f({{\bf{p}}_{AB}}|{{\bf{D}}_{{A_1}{B_1}}},{{\bf{D}}_{{A_2}{B_2}}})\nonumber\\
& =& \displaystyle \frac{{f({{\bf{D}}_{{A_2}{B_2}}}|{\bf{p}}_{{A_1}{B_1}})f({\bf{p}}_{{A_1}{B_1}})}}{{f({{\bf{D}}_{{A_2}{B_2}}})}}\nonumber\\
 &=& \displaystyle \frac{{f({{\bf{D}}_{{A_2}{B_2}}}|{\bf{p}}_{{A_1}{B_1}})f({\bf{p}}_{{A_1}{B_1}})}}{{\displaystyle \int {f({{\bf{D}}_{{A_2}{B_2}}}|{\bf{p}}_{{A_1}{B_1}})f({\bf{p}}_{{A_1}{B_1}})} d{\bf{p}}_{{A_1}{B_1}}}}.
\label{comb_bayes}
\end{eqnarray}
In the equation, ${\bf{p}}_{{A_1}{B_1}}$ is derived from  ${{\bf{D}}_{{A_1}{B_1}}}$ that follows Dirichlet distribution, so its pdf can be computed as follows. 
\begin{eqnarray}
&&f({\bf{p}}_{{A_1}{B_1}}) \nonumber\\
 &=& \frac{{\Gamma ({\alpha _{{A_1}{B_1}}} + {\beta _{{A_1}{B_1}}} + {\gamma _{{A_1}{B_1}}})}}{{\Gamma ({\alpha _{{A_1}{B_1}}})\Gamma ({\beta _{{A_1}{B_1}}})\Gamma ({\gamma _{{A_1}{B_1}}})}}\times \nonumber\\
 &&({p_1})^{{\alpha _{{A_1}{B_1}}} - 1}{({p_2})^{{\beta _{{A_1}{B_1}}} - 1}}({p_3})^{{\gamma _{{A_1}{B_1}}} - 1}.\nonumber\\
\label{comb_bayes_1}
\end{eqnarray}

On the other hand, because ${{\bf{D}}_{{A_2}{B_2}}}$ follows the multinomial distribution, derived from ${\bf{p}}_{{A_1}{B_1}}$, its pdf can be expressed as
\begin{eqnarray}
&&f({{\bf{D}}_{{A_2}{B_2}}}|{\bf{p}}_{{A_1}{B_1}}) \nonumber\\
 &=& \frac{{\Gamma ({\alpha _{{A_2}{B_2}}} + {\beta _{{A_2}{B_2}}} + {\gamma _{{A_2}{B_2}}} + 1)}}{{\Gamma ({\alpha _{{A_2}{B_2}}} + 1)\Gamma ({\beta _{{A_2}{B_2}}} + 1)\Gamma ({\gamma _{{A_2}{B_2}}} + 1)}}\times \nonumber\\
&& {({p_1})^{{\alpha _{{A_2}{B_2}}}}}{({p_2})^{{\beta _{{A_2}{B_2}}}}}{({p_3})^{{\gamma _{{A_2}{B_2}}}}}.\nonumber\\
\label{comb_bayes_2}
\end{eqnarray}
Substituting Eq.~\ref{comb_bayes_1} and Eq.~\ref{comb_bayes_2} for $f({\bf{p}}_{{A_1}{B_1}})$ and $f({{\bf{D}}_{{A_2}{B_2}}}|{\bf{p}}_{{A_1}{B_1}})$ in Eq.~\ref{comb_bayes}, the analytic form of Eq.~\ref{comb_evid} will be
\begin{eqnarray}
&&f({{\bf{p}}_{AB}}|{{\bf{D}}_{{A_1}{B_1}}},{{\bf{D}}_{{A_2}{B_2}}}) \nonumber\\
&=& \frac{{\Gamma ({\alpha _{AB}} + {\beta _{AB}} + {\gamma _{AB}})}}{{\Gamma ({\alpha _{AB}})\Gamma ({\beta _{AB}})\Gamma ({\gamma _{AB}})}} \times \nonumber\\
&&{({p_1})^{{\alpha _{AB}} - 1}}{({p_2})^{{\beta _{AB}} - 1}}{({p_3})^{{\gamma _{AB}} - 1}},\nonumber\\
\label{comb_ana}
\end{eqnarray} 
where
\begin{eqnarray}
 \alpha _{AB} &=&   \alpha _{{A_1}{B_1}}^{} + \alpha _{{A_2}{B_2}}^{} ,\nonumber \\
 \beta _{AB} &=& \beta _{{A_1}{B_1}}^{} + \beta _{{A_2}{B_2}}^{},\nonumber \\
\gamma _{AB}  &=& \gamma _{{A_1}{B_1}}^{} + \gamma _{{A_2}{B_2}}^{}. \nonumber
\end{eqnarray}
Obviously,  Eq.~\ref{comb_ana} can be considered the most likely pdf of the following Dirichlet distribution.
\[Dir({\alpha _{{A_1}{B_1}}} + {\alpha _{{A_2}{B_2}}},{\beta _{{A_1}{B_1}}} + {\beta _{{A_2}{B_2}}},{\gamma _{{A_1}{B_1}}} + {\gamma _{{A_2}{B_2}}}).\]
Then, the equation
\begin{eqnarray}
\int {f(x|{{\bf{p}}_{AB}})f({{\bf{p}}_{AB}}|{{\bf{D}}_{{A_1}{B_1}}}, {{\bf{D}}_{{A_2}{B_2}}})d({{\bf{p}}_{AB}})}\nonumber
\end{eqnarray}
can be regarded as a DC distribution upon observations $\{ {\alpha _{{A_1}{B_1}}} + {\alpha _{{A_2}{B_2}}}, {\beta _{{A_1}{B_1}}} + {\beta _{{A_2}{B_2}}}, {\gamma _{{A_1}{B_1}}}  + {\gamma _{{A_2}{B_2}}} \}$.
The above description essentially reflects the important property of DC distribution: two DC distributions can be combined, by adding up the corresponding controlling hyper-parameters, to yield a new DC distribution.

According to the definition of an opinion, the analytic form of Eq.~\ref{Com_Opr_b} can be expressed as 
 \begin{eqnarray}
&&\int {f(x = 1|{{\bf{p}}_{AB}})f({{\bf{p}}_{AB}}|{{\bf{D}}_{{A_1}{B_1}}},{{\bf{D}}_{{A_2}{B_2}}})d({{\bf{p}}_{AB}})} \nonumber\\
 &=& \frac{{{\alpha _{{A_1}{B_1}}} + {\alpha _{{A_2}{B_2}}}}}{{{\alpha _{{A_1}{B_1}}} + {\alpha _{{A_2}{B_2}}} + {\beta _{{A_1}{B_1}}} + {\beta _{{A_2}{B_2}}} + {\gamma _{{A_1}{B_1}}} + {\gamma _{{A_2}{B_2}}}}}\nonumber.
\end{eqnarray}
The probability that $B$ will not behave as what $A$ expects and the probability that $B$ will behave in an ambiguous way can be expressed as
 \begin{eqnarray}
&&\int {f(x =2|{{\bf{p}}_{AB}})f({{\bf{p}}_{AB}}|{{\bf{D}}_{{A_1}{B_1}}},{{\bf{D}}_{{A_2}{B_2}}})d({{\bf{p}}_{AB}})} \nonumber\\
 &=& \frac{{{\beta _{{A_1}{B_1}}} + {\beta _{{A_2}{B_2}}}}}{{{\alpha _{{A_1}{B_1}}} + {\alpha _{{A_2}{B_2}}} + {\beta _{{A_1}{B_1}}} + {\beta _{{A_2}{B_2}}} + {\gamma _{{A_1}{B_1}}} + {\gamma _{{A_2}{B_2}}}}}\nonumber,
\end{eqnarray}
and
 \begin{eqnarray}
&&\int {f(x =3|{{\bf{p}}_{AB}})f({{\bf{p}}_{AB}}|{{\bf{D}}_{{A_1}{B_1}}},{{\bf{D}}_{{A_2}{B_2}}})d({{\bf{p}}_{AB}})} \nonumber\\
 &=& \frac{{{\gamma _{{A_1}{B_1}}} + {\gamma_{{A_2}{B_2}}}}}{{{\alpha _{{A_1}{B_1}}} + {\alpha _{{A_2}{B_2}}} + {\beta _{{A_1}{B_1}}} + {\beta _{{A_2}{B_2}}} + {\gamma _{{A_1}{B_1}}} + {\gamma _{{A_2}{B_2}}}}}\nonumber,
\end{eqnarray}
respectively.
As such, we are able to formally define the combining operation as follows.
\begin{definition}[Combining Operation]
Let $\omega _{A_1B_1} = \left<\alpha_{A_1B_1}, \beta_{A_1B_1}, \gamma_{A_1B_1}\right>$ and $\omega _{A_2B_2} = \left<\alpha_{A_2B_2}, \beta_{A_2B_2}, \gamma_{A_2B_2}\right>$ be the two opinions $A$ has on $B$, the combining operation $\Theta (\omega _{A_1B_1},\omega _{A_2B_2})$ is carried out as follows.
\begin{equation}
\Theta(\omega _{A_1B_1},\omega _{A_2B_2}) = \left<\alpha _{AB}, \beta _{AB}, \gamma _{AB}\right>,
\end{equation}
where
\begin{equation}
\left\{
\begin{array}{l}
\alpha _{AB}= \displaystyle {{\alpha _{{A_1}{B_1}}} + {\alpha _{{A_2}{B_2}}}}\\
\beta _{AB} = \displaystyle {{\beta _{{A_1}{B_1}}} + {\beta _{{A_2}{B_2}}}}\\
\gamma _{AB} = \displaystyle {{\gamma _{{A_1}{B_1}}} + {\gamma _{{A_2}{B_2}}}}
\end{array}.
\right.
\label{eq:combining}
\end{equation}
\label{t1}
\end{definition}

It is worth mentioning that the combining operation yields two properties: commutative and associative proprieties.
\begin{corollary}
Commutative Property:
Given two independent opinions $\omega _{{A_1}{B_1}}$ and $\omega _{{A_2}{B_2}}$, $\Theta ({\omega _{{A_1}{B_1}}},{\omega _{{A_2}{B_2}}}) \equiv \Theta ({\omega _{{A_2}{B_2}}},{\omega _{{A_1}{B_1}}})$.
\label{c1}
\end{corollary}

\begin{proof}
Based on Eq.~\ref{eq:combining}.
\end{proof}

\begin{corollary}
Associative Property:
Given three independent opinions $\omega _{{A_1}{B_1}}$, $\omega _{{A_2}{B_2}}$ and $\omega _{{A_3}{B_3}}$, then $\Theta ({\omega _{{A_1}{B_1}}},\Theta ({\omega _{{A_2}{B_2}}},{\omega _{{A_3}{B_3}}})) \equiv \Theta (\Theta ({\omega _{{A_1}{B_1}}},{\omega _{{A_2}{B_2}}}),{\omega _{{A_3}{B_3}}})$.
\label{c2}
\end{corollary}

\begin{proof}
Based on Eq.~\ref{eq:combining}.
\end{proof}

If $A$ has more than two opinions on $B$, e.g., $\omega _{{A_1}{B_1}}, \omega _{{A_2}{B_2}} \cdots \omega _{{A_n}{B_n}}$, these opinion can be combined by $\Theta (\Theta (\Theta (\omega _{{A_1}{B_1}},\omega _{{A_2}{B_2}}), \cdots),\omega _{{A_n}{B_n}})$. 
As combining operation is commutative and associative, it can be rewritten as $\Theta (\omega _{{A_1}{B_1}},\omega _{{A_2}{B_2}}, \cdots \omega _{{A_n}{B_n}})$. 

\subsection{Expected Belief of An Opinion}
\label{CH:ExpB}
With the proposed discounting and combining operations, the trust between two users in an OSN can be computed, which will be elaborated in details in Section \ref{CH:algorithm}.
Note that the computed trust is in the form of an opinion.
To transform an opinion into a trust value, i.e., the probability that a user is trustworthy, we need to design a mapping mechanism.

Given an opinion $\omega _{AX} = \left< {\alpha _{AX}},{\beta _{AX}},{\gamma _{AX}}\right>$, it is of interest to know how likely $X$ will perform the desired action(s) requested by $A$.
We call this probability as the expected belief of $\omega_{AX}$. 
Although $\alpha _{AX}$ denotes the belief of opinion $\omega_{AX}$, components $\beta _{AX}$, $\gamma _{AX}$ also need to be considered in computing the expected belief.

We know that $\alpha_{AX}$ and $\beta _{AX}$ are the numbers of (negative and positive) certain evidence, so they must be used in computing the expected belief. 
$\gamma _{AX}$ only records the uncertain evidence, so it should be omitted in the computation of expected belief.
Ignoring uncertain evidence, DC distribution of $\omega_X^A$ is collapsed into a Beta-Categorical (BC) distribution.
\begin{eqnarray}
\label{ecbeta}
&& f({p_1},{p_2}\left| {\alpha_{AX} ,\beta_{AX} } \right.) \nonumber\\
&=& \frac{{\Gamma (\alpha_{AX}  +\beta_{AX} )}}{{\Gamma (\alpha_{AX}) \cdot \Gamma (\beta_{AX} )}}\cdot {(1 - {p_1})^{\alpha_{AX}  - 1}}p_2^{\beta_{AX}  - 1} .\nonumber
\end{eqnarray}
Consequently, the original opinion is collapsed into
\begin{equation}
\omega_{AX} = \left<\alpha_{AX},\beta_{AX} \right>.\nonumber\\
\end{equation}

With the collapsed opinion, we apply the approach proposed in~\cite{wang2007formal} to compute the expected belief as follows.
\begin{eqnarray}
E_{{\omega _{AX}}} &=& \left(\frac{{{\alpha _{AX}}}}{{{\alpha _{AX}} + {\beta _{AX}}}} + \frac{{{\beta _{AX}}}}{{{\alpha _{AX}} + {\beta _{AX}}}} \right)a_{AX} \nonumber\\
 &\times& (1 - c_{AX}) + \frac{{{\alpha _{AX}}}}{{{\alpha _{AX}} + {\beta _{AX}}}} \cdot c_{AX}\nonumber\\
 &=& \frac{{{\alpha _{AX}}}}{{{\alpha _{AX}} + {\beta _{AX}}}} \cdot c_{AX} + a_{AX} \cdot (1 - c_{AX}), \nonumber\\
 \label{e9}
\end{eqnarray}
where $c_{AX}$ is the certainty factor~\cite{wang2007formal} of a Beta distribution, and $a_{AX}$ is the base rate. The certainty factor $c_{AX}$, ranging from $0$ to $1$, is determined by the total amount of certain evidence and the ratio between positive and negative evidence. 
\begin{equation}
\label{e_cer}
c_{AX} = \frac{1}{2}\int_0^1 {\left| {\frac{1}{{B(\alpha_{AX},\beta_{AX})}}{x^{\alpha_{AX}}}(1 - {x^{\beta_{AX}}}) - 1} \right|} dx.
\end{equation}
Basically, $c_{AX}$ approaches to $1$ when the amount of certain evidence or the disparity between positive and negative evidence is large. 

\section{AssessTrust Algorithm}
\label{CH:algorithm}
Based on 3VSL and the discounting and combining operations, we design the AssessTrust (AT) algorithm to conduct trust assessment in social networks with arbitrary topologies.
Here, we treat a social network as a two-terminal directed graph (TTDG), in which the two terminals represent the trustor and trustee, respectively.
Obviously, the trustor and trustee must be different users because a trustor will never evaluate the trust of itself. 
As a TTDG is not necessarily a directed acyclic graph, there may be cycles in the network.

To ensure AT works in arbitrary topologies, we need to first prove AT can handle non-series-parallel network topologies.
This is a challenge because the only operations available for trust computation are the discounting and combining operations. 
The discounting/combining operation requires that the network topologies must be series/parallel. 
We address this challenge by differentiating distorting opinions from original opinions in trust propagation. 
For example, if $A$ trusts $B$ and $B$ trusts $C$, then $A$'s opinion on $B$ is called the distorting opinion, and $B$'s opinion on $C$ is the original opinion. 
We discover that, in trust fusion, the original opinions can be used only once but the distorting opinions can be used any number of times. 
This is because the distorting opinion only depreciates certain evidence into uncertain evidence, \ie, it does not change the total amount of evidence.
On the other hand, when two (discounted) original opinions are combined, the total number of evidence in the resulting opinion will be increased.

In addition, we have to further show that AT works in arbitrary TTDGs. 
This is a challenge because it is impossible to test AT in all possible network topologies.
We address this challenge by mathematically proving that AT works in arbitrary networks. 
After addressing these two challenges, we present the AT algorithm and use an example to illustrate how is works.

\subsection{Properties of Different Opinions}
\label{sec:op}
For the two opinions involved in a discounting operation, their functionality are different, regarding to trust computation in an OSN.
\begin{definition}[Distorting and Original Opinions]
Given a discounting operation $\Delta(\omega_{AB}, \omega_{BC}) $, we define $\omega_{AB}$ as the distorting opinion, and $\omega_{BC}$ the original opinion.
\end{definition}
\begin{figure}
\centering
\subfigure[]{\includegraphics[width=1.5in]{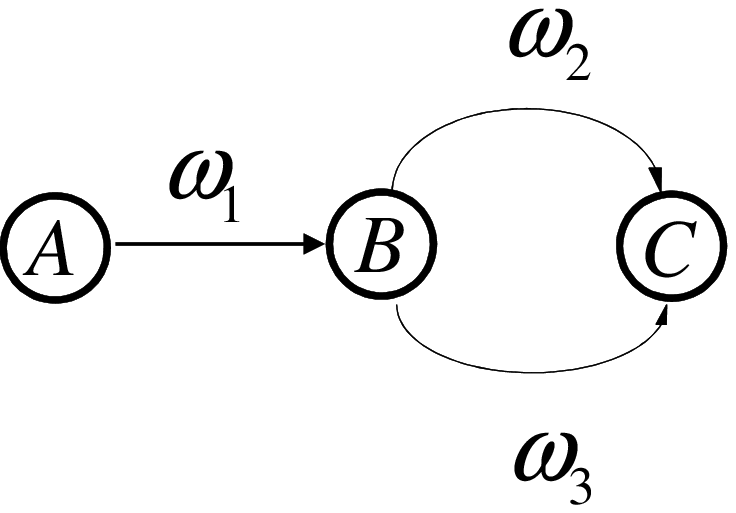}
\label{DO1}}
\hfil
\subfigure[]{\includegraphics[width=1.5in]{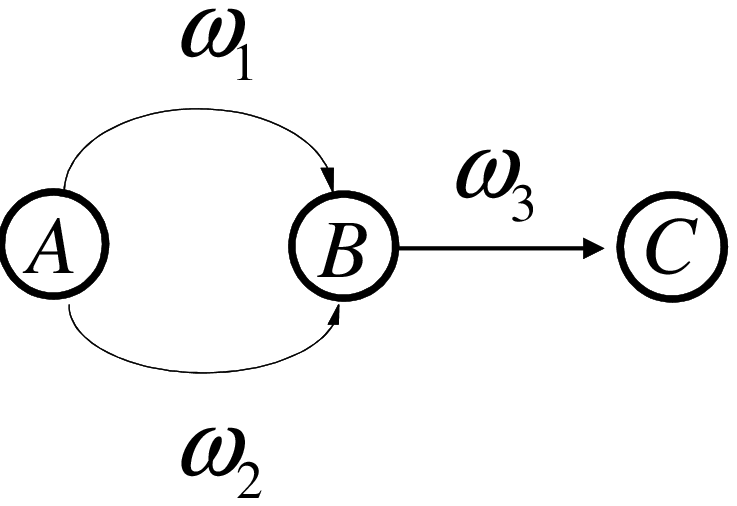}
\label{DO2}}
\caption{Difference between distorting and original opinions}
\label{different operation}
\end{figure}

To understand the difference between the distorting and original opinions, we study two special cases, as shown in Fig.~\ref{different operation}. 
The detailed study reveals that a distorting opinion can be used several times in trust computation but an original opinion can be used only once.
\begin{theorem}
\label{thm:1}
Let $\omega _{B_1C_1} = \left<\alpha_{B_1C_1}, \beta_{B_1C_1}, \gamma_{B_1C_1}\right>$ and $\omega _{B_2C_2} = \left<\alpha_{B_2C_2}, \beta_{B_2C_2}, \gamma_{B_2C_2}\right>$ be two opinions $B$ has on $C$. 
Let $\omega _{AB} = (\alpha_{AB}, \beta_{AB}, \gamma_{AB})$ be $A$'s opinion on $B$, then we always have
\begin{eqnarray}
&&\Theta (\Delta ({\omega _{AB}},{\omega _{{B_1}{C_1}}}),\Delta ({\omega _{AB}},{\omega _{{B_2}{C_2}}})) \nonumber\\
&\equiv & \Delta ({\omega _{AB}},\Theta ({\omega _{{B_1}{C_1}}},{\omega _{{B_2}{C_2}}})).
\label{distributive}
\end{eqnarray}
\label{t_distributive}
\label{l10}
\end{theorem}
\begin{proof}
Let's take a look at the left side of Eq.~\ref{distributive}.
According to the definition of discounting operation, the result of $\Delta ({\omega _{AB}},{\omega _{{B_1}{C_1}}})$ can be written as
\begin{eqnarray}
{\omega _{{A}C_1}} &=& \Delta ({\omega _{AB}},{\omega _{{B_1}C_1}})\nonumber\\
 &=& \left<{\alpha _{{A}C_1}},{\beta _{{A}C_1}},{\gamma _{{A}C_1}}\right>,\nonumber
\end{eqnarray}
where
\begin{eqnarray}
{\alpha _{A{C_1}}} &=& \frac{{{\alpha _{AB}}{\alpha _{{B_1}{C_1}}}}}{{{\alpha _{AB}} + {\beta _{AB}} + {\gamma _{AB}}}},\nonumber\\
{\beta _{A{C_1}}} &=& \frac{{{\alpha _{AB}}{\beta _{{B_1}{C_1}}}}}{{{\alpha _{AB}} + {\beta _{AB}} + {\gamma _{AB}}}},\nonumber\\
{\gamma _{A{C_1}}} &=& \frac{{({\beta _{AB}} + {\gamma _{AB}})({\alpha _{{B_1}{C_1}}} + {\beta _{{B_1}{C_1}}} + {\gamma _{{B_1}{C_1}}})}}{{{\alpha _{AB}} + {\beta _{AB}} + {\gamma _{AB}}}}\nonumber\\
 &+& \frac{{{\alpha _{AB}}{\gamma _{{B_1}{C_1}}}}}{{{\alpha _{AB}} + {\beta _{AB}} + {\gamma _{AB}}}}.\nonumber\\
\end{eqnarray}
The result of $\Delta ({\omega _{AB}},{\omega _{{B_2}{C_2}}})$ can be written as
\begin{eqnarray}
{\omega _{A{C_2}}} &=& \Delta ({\omega _{AB}},{\omega _{{B_2}{C_2}}})\nonumber\\
 &=& \left<{\alpha _{A{C_2}}},{\beta _{A{C_2}}},{\gamma _{A{C_2}}}\right>,\nonumber
\end{eqnarray}
where
\begin{eqnarray}
{\alpha _{A{C_2}}} &=& \frac{{{\alpha _{AB}}{\alpha _{{B_2}{C_2}}}}}{{{\alpha _{AB}} + {\beta _{AB}} + {\gamma _{AB}}}},\nonumber\\
{\beta _{A{C_2}}} &=& \frac{{{\alpha _{AB}}{\beta _{{B_2}{C_2}}}}}{{{\alpha _{AB}} + {\beta _{AB}} + {\gamma _{AB}}}},\nonumber\\
{\gamma _{A{C_2}}} &=& \frac{{({\beta _{AB}} + {\gamma _{AB}})({\alpha _{{B_2}{C_2}}} + {\beta _{{B_2}{C_2}}} + {\gamma _{{B_2}{C_2}}})}}{{{\alpha _{AB}} + {\beta _{AB}} + {\gamma _{AB}}}}\nonumber\\
 &+& \frac{{{\alpha _{AB}}{\gamma _{{B_2}{C_2}}}}}{{{\alpha _{AB}} + {\beta _{AB}} + {\gamma _{AB}}}}.
\end{eqnarray}

If these two opinions are combined, we will have
\begin{eqnarray}
{\omega _{AC}} &=& \Theta (\Delta ({\omega _{{A_1}{B_1}}},{\omega _{BC}}),\Delta ({\omega _{{A_2}{B_2}}},{\omega _{BC}})) \nonumber\\
&=& \left<{\alpha _{AC}},{\beta _{AC}},{\gamma _{AC}}\right>, \nonumber
\end{eqnarray}
where
\begin{eqnarray}
{\alpha _{AC}} &=& \frac{{{\alpha _{AB}}{\alpha _{{B_1}{C_1}}} + {\alpha _{AB}}{\alpha _{{B_2}{C_2}}}}}{{{\alpha _{AB}} + {\beta _{AB}} + {\gamma _{AB}}}},\nonumber\\
{\beta _{AC}} &=& \frac{{{\alpha _{AB}}{\beta _{{B_1}{C_1}}} + {\alpha _{AB}}{\beta _{{B_2}{C_2}}}}}{{{\alpha _{AB}} + {\beta _{AB}} + {\gamma _{AB}}}},\nonumber
\end{eqnarray}
\begin{eqnarray}
{\gamma _{AC}} &=& \frac{{({\beta _{AB}} + {\gamma _{AB}})({\alpha _{{B_1}{C_1}}} + {\beta _{{B_1}{C_1}}} + {\gamma _{{B_1}{C_1}}})}}{{{\alpha _{AB}} + {\beta _{AB}} + {\gamma _{AB}}}}\nonumber\\
 &+& \frac{{{\alpha _{AB}}{\gamma _{{B_1}{C_1}}}}}{{{\alpha _{AB}} + {\beta _{AB}} + {\gamma _{AB}}}}\nonumber\\
 &+& \frac{{({\beta _{AB}} + {\gamma _{AB}})({\alpha _{{B_2}{C_2}}} + {\beta _{{B_2}{C_2}}} + {\gamma _{{B_2}{C_2}}})}}{{{\alpha _{AB}} + {\beta _{AB}} + {\gamma _{AB}}}}\nonumber\\
 &+& \frac{{{\alpha _{AB}}{\gamma _{{B_2}{C_2}}}}}{{{\alpha _{AB}} + {\beta _{AB}} + {\gamma _{AB}}}}.\nonumber
\end{eqnarray}
Now, we look at the right side of Eq.~\ref{distributive}.
The term $\Theta ({\omega _{{B_1}{C_1}}},{\omega _{{B_2}{C_2}}})$ can be written as
\begin{eqnarray}
{\omega _{BC}} &=& \Theta ({\omega _{{B_1}{C_1}}},{\omega _{{B_2}{C_2}}})\nonumber \\
&=& \left<\alpha _{BC},\beta _{BC},\gamma _{BC}\right>,
\end{eqnarray}
where
\begin{eqnarray}
{\alpha _{BC}} &=& {\alpha _{{B_1}{C_1}}} + {\alpha _{{B_2}{C_2}}},\nonumber\\
{\beta _{BC}} &=& {\beta _{{B_1}{C_1}}} + {\beta _{{B_2}{C_2}}},\nonumber\\
{\gamma _{BC}} &=& {\gamma _{{B_1}{C_1}}} + {\gamma _{{B_2}{C_2}}}.\nonumber
 \label{sub_comb}
\end{eqnarray}
Putting Eq.~\ref{sub_comb} back into the equation, we have
\begin{eqnarray}
\omega_{AC}^{\prime}&=&\Delta ({\omega _{AB}},\Theta ({\omega _{{B_1}{C_1}}},{\omega _{{B_2}{C_2}}}))\nonumber\\
 &= &\left<{\alpha _{AC}^{\prime}},{\beta _{AC}^{\prime}},{\gamma _{AC}^{\prime}}\right>,\nonumber
\end{eqnarray}
where
\begin{eqnarray}
{\alpha _{AC}^{\prime}} &=& \frac{{{\alpha _{AB}}({\alpha _{{B_1}{C_1}}} + {\alpha _{{B_2}{C_2}}})}}{{{\alpha _{AB}} + {\beta _{AB}} + {\gamma _{AB}}}} \nonumber\\
&=& \frac{{{\alpha _{AB}}{\alpha _{{B_1}{C_1}}} + {\alpha _{AB}}{\alpha _{{B_2}{C_2}}}}}{{{\alpha _{AB}} + {\beta _{AB}} + {\gamma _{AB}}}},\nonumber\\
{\beta _{AC}^{\prime}} &=& \frac{{{\alpha _{AB}}({\beta _{{B_1}{C_1}}} + {\beta _{{B_2}{C_2}}})}}{{{\alpha _{AB}} + {\beta _{AB}} + {\gamma _{AB}}}} \nonumber\\
&=& \frac{{{\alpha _{AB}}{\beta _{{B_1}{C_1}}} + {\alpha _{AB}}{\beta _{{B_2}{C_2}}}}}{{{\alpha _{AB}} + {\beta _{AB}} + {\gamma _{AB}}}},\nonumber\\
{\gamma _{AC}^{\prime}} &=& \frac{{({\beta _{AB}} + {\gamma _{AB}})({\alpha _{{B_1}{C_1}}} + {\beta _{{B_1}{C_1}}} + {\gamma _{{B_1}{C_1}}})}}{{{\alpha _{AB}} + {\beta _{AB}} + {\gamma _{AB}}}}\nonumber\\
 &+& \frac{{{\alpha _{AB}}{\gamma _{{B_1}{C_1}}}}}{{{\alpha _{AB}} + {\beta _{AB}} + {\gamma _{AB}}}}\nonumber\\
 &+& \frac{{({\beta _{AB}} + {\gamma _{AB}})({\alpha _{{B_2}{C_2}}} + {\beta _{{B_2}{C_2}}} + {\gamma _{{B_2}{C_2}}})}}{{{\alpha _{AB}} + {\beta _{AB}} + {\gamma _{AB}}}}\nonumber\\
 &+& \frac{{{\alpha _{AB}}{\gamma _{{B_2}{C_2}}}}}{{{\alpha _{AB}} + {\beta _{AB}} + {\gamma _{AB}}}}.
\end{eqnarray}
Clearly, $\omega _{AC}^{\prime}$ is equivalent to $\omega _{AC}$.
\end{proof}

\begin{theorem}
\label{thm:1-1}
Let $\omega _{A_1B_1} = (\alpha_{A_1B_1}, \beta_{A_1B_1}, \gamma_{A_1B_1})$ and $\omega _{A_2B_2} = (\alpha_{A_2B_2}, \beta_{A_2B_2}, \gamma_{A_2B_2})$ be $A$'s two opinions on $B$.
Let $\omega _{BC} = (\alpha_{BC}, \beta_{BC}, \gamma_{BC})$ be $B$'s opinion on $C$, then the following equation \textbf{does not} hold.
\begin{eqnarray}
&&\Theta (\Delta ({\omega _{A_{1}B_{1}}},{\omega _{{B}{C}}}),\Delta ({\omega _{A_{2}B_{2}}},{\omega _{{B}{C}}})) \nonumber\\
&\equiv & \Delta (\Theta ({\omega _{{A_1}{B_1}}},{\omega _{{A_2}{B_2}}}), {\omega _{BC}}).
\label{l101}
\end{eqnarray}
\label{l1011}
\end{theorem}

\begin{proof}
In Section~\ref{CH:3vsl}, we have shown that the combining operation can be applied in $\Theta(\omega_{{A_1}{B_1}}, \omega_{{A_2}{B_2}})$ only if the evidence in $\omega_{{A_1}{B_1}}$ and $\omega_{{A_2}{B_2}}$ are independent.
In the left side of Eq.~\ref{l101}, opinions $\Delta (\omega_{{A_1}{B_1}},\omega_{BC})$ and $\Delta (\omega_{{A_2}{B_2}},\omega_{BC})$ share the same evidence from the opinion $\omega_{BC}$. 
As a result, the combining operation does not apply here. 
Therefore, $\Delta (\Theta(\omega_{{A_1}{B_1}},\omega_{{A_2}{B_2}}),\omega_{BC})$ is the only correct solution, and it does not equal to $\Theta(\Delta (\omega_{{A_1}{B_1}},\omega_{BC}),\Delta (\omega_{{A_2}{B_2}},\omega_{BC}))$.
\end{proof}

From Theorems~\ref{l10} and~\ref{l1011}, we note that reusing $\omega_{{A}{B}}$ in case (a) is allowed but reusing $\omega_{BC}$ in case (b) is not.

The difference between $\omega_{{A}{B}}$ and $\omega_{BC}$ is that $\omega_{{A}{B}}$ is a distorting opinion while $\omega_{BC}$ is an original opinion. 
Therefore, we conclude that in trust computation, an original opinion can be combined only once, while a distorting opinion can be used any number of times, because it does not change the total amount of evidence in the resulting opinion.

\subsection{Arbitrary Network Topology}
\label{sec:alg}
As the distorting and original opinions are distinguished, we will prove that 3VSL is capable of handling non-series-parallel network topologies.

\begin{theorem}
Given an arbitrary two-terminal directed graph $G=(V,E)$ where $A$, $C$ are the first and second terminals, or the trustor and trustee. 
In the graph, a vertex $u$ represents a user, the edge $e(u,v)$ denotes $u$'s opinion about $v$'s trust, denoted as $\omega_{uv}$. By applying the discounting and combining operations, the resulting opinion $\omega_{AC}$ is solvable and unique.
\label{t2}
\end{theorem}

\begin{proof}
We prove the theorem in a recursive manner, \ie, reducing the original problem into sub-problem(s) and continuing to reduce the sub-problems until the base case is solvable and yields a unique solution.
\begin{figure}[!t]
\centering
\includegraphics[width=2.5in]{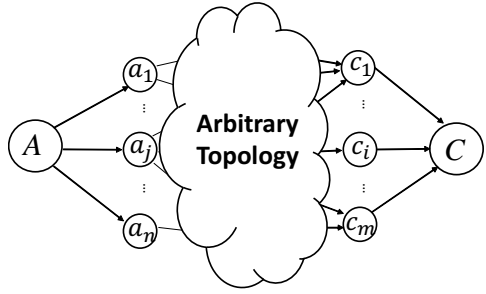}
\caption{Illustration of an arbitrary network topology}
\label{arbitrary}
\end{figure}

As shown in Fig.~\ref{arbitrary}, we assume there are $m$ nodes ($c_1, c_2, \cdots, c_m$) connecting to $C$, \ie, $e(c_i,C) \in E$ where $i = 1, 2, \cdots, m$. There are $n$ nodes ($a_1, a_2, \cdots, a_n$) being connected from $A$, \ie, $e(A,a_j) \in E$ where $j = 1, 2, \cdots, n$.
\begin{flushleft}
\underline{Reduction rules}
\end{flushleft}
\noindent \textit{Case 1}: If there is only one node connecting to $C$, \ie, $m=1$, then $\omega_{AC} = \Delta (\omega_{A{c_1}}, \omega_{{c_1}C}) $. In this case, we reduce the problem of computing $\omega_{AC}$ to calculating $\omega_{A{c_1}}$.
We know $A$ and $c_1$ are connected in a smaller sub-graph.

\noindent \textit{Case 2}: If there is more than one node connected to $C$, \ie, $m>1$, $\omega_{AC}$ is equal to $\Theta ( \Delta (\omega_{A{c_1}}, \omega_{{c_1}C}), \Delta (\omega_{A{c_2}}, \omega_{{c_2}C}), \cdots, \Delta (\omega_{A{c_m}}, \omega_{{c_m}C}))$ due to Theorem~\ref{l10}. Therefore, $\omega_{AC}$ is solvable and unique if and only if each $\omega_{A{c_i}}$ is solvable and unique, where $\omega_{A{c_i}}$ can be obtained from the sub-graph $G'$, in which edges $\{ e(c_i, C)\}$ and node $C$ are removed from $G$. In this case, we reduce the problem of computing $\omega_{AC}$ to computing $\omega_{A{c_i}}$.

In each round of reduction, $G$ is reduced into a smaller graph, with $|E| = |E|- m $ and $|V| = |V| - 1$. After applying the reduction rules on sub-problems recursively, the base case will be eventually reached, \ie, $|E| = 1$ and $|V| = 2$.
\begin{flushleft}
\underline{Base Case}
\end{flushleft}
\indent The graph of base case contains only one edge from $A$ to $a_j$ where $j = 1, 2, \cdots, n$. 
As $\omega_{A{a_j}}$ is known, the base case is solvable and its solution is unique. Applying the equations in Case 1 and 2 repeatedly, we can obtain a unique solution to $\omega_{AC}$.
\end{proof}

\subsection{Differences between 3VSL and SL}
\label{advtg_3vsl}
The major difference between SL and 3VSL lies in the definition of uncertainty in the trust models.
In 3VSL, the uncertainty in a trust opinion is measured by the number of uncertain evidence.
However, the amount of uncertain evidence in a SL opinion is always 2.
Because uncertain evidence is obtained if an ambiguous behavior of a trustee is observed, it could not be a constant number. 

We take an example to explain the different definitions of uncertainty in SL and 3VSL models.
Let's consider a series topology composed of $A$, $B$ and $C$, as shown in Fig~\ref{s-2}.
We assume opinions $\omega_{AB} \left\langle {5,3,2} \right\rangle$ and $\omega_{BC} = \left\langle {4,4,2} \right\rangle$.
Then, $A$'s opinion of $C$'s trust can be computed by applying the discounting operation, defined in SL or 3VSL, on opinions $\omega_{AB}$ and $\omega_{BC}$, \ie, $\omega_{AC} = \Delta (\omega_{AB}, \omega_{BC})$.
With the SL model,  we have $\omega_{AC} = \left\langle {{2 \mathord{\left/
 {\vphantom {2 3}} \right.
 \kern-\nulldelimiterspace} 3},{2 \mathord{\left/
 {\vphantom {2 3}} \right.
 \kern-\nulldelimiterspace} 3},2} \right\rangle$.
Apparently, $10/3$ positive evidence and $10/3$ negative evidence are removed from the original evidence space.
In other words, the amount of certain evidence shrinks for $83\%$, \ie, $83\%$ of evidence are distorted and disappear.
Based on the SL model, we know the belief component $b_{AB}$ in opinion $\omega_{AB}$ equals to $5/(5+3+2) = 0.5$, \ie, with $50\%$ of chance, $A$ could trust $B$'s recommendation.
That also implies only $50\%$ of evidence should be distorted from $B$'s opinion of $C$, which is not the case in the example.

In contrast, 3VSL model introduces an uncertainty state to keep tracking of the uncertain evidence generated when trust propagates within an OSN. 
In 3VSL, we have $\omega_{AC} = \left\langle {2,2,6} \right\rangle$.
The total number of evidence in the resulting opinion $\omega_{AC}$ is the same as $\omega_{BC}$, \ie, $\alpha_{AC} + \beta_{AC} + \gamma_{AC} = \alpha_{BC} + \beta_{BC} + \gamma_{BC} = 10$.
In fact, only $50\%$ of certain evidence from $\alpha_{BC}$ and $\beta_{BC}$ are transferred into $\gamma_{AC}$.
%
Clearly, 3VSL leverages the uncertainty state to store the ``distorted'' positive and negative evidence in trust propagation and hence achieves better accuracy. 
This hypothesis will be validated in Section~\ref{CH:eval}.

Another difference is that 3VSL is capable to handle a social network with arbitrary topologies while SL cannot.
It is well-known that SL can only handle series-parallel network topologies. 
A series-parallel graph can be decomposed into many series (see Fig.~\ref{series_topo}) or parallel (see Fig.~\ref{parallel_topo}) sub-graphs so that every edge in the original graph will appear only once in the sub-graphs \cite{jakoby2001space}.
In real-world social networks, however, the connection between two users could be too complicated to be decomposed into series-parallel graphs.
To apply the SL model, a complex topology has to be simplified into a series-parallel topology by removing or selecting edges~\cite{hang2009operators, hang2010trust,Hang:2009:OPT:1558109.1558155}.
However, it is not clear which edges need to be removed in a large-scale OSN.
As a result, the solutions proposed in~\cite{hang2009operators, hang2010trust,Hang:2009:OPT:1558109.1558155} cannot be implemented.
In 3VSL, the difference between distorting and original opinions is first identified, and then a recursive algorithm is designed accordingly.
The algorithm is able to process social networks with complex topologies, even with cycles.

\subsection{AssessTrust Algorithm}
\begin{algorithm}[ht]
\label{A1}
\caption{AssessTrust($G$, $A$, $C$, $H$)}
\begin{algorithmic}[1]
\REQUIRE $G$, $A$, $C$, and $H$.
\ENSURE  $\Omega_{AC}$.
\STATE $n \gets 0$
\IF {$H > 0$}
\FORALL  {incoming edges $e(c_i,C)$ $\in$ $G$}
\IF {$c_i = A$}
\STATE $\Omega_i \gets \omega_{{c_i}C}$
\ELSE
\STATE $G' \gets G - e({c_i},C) $
\STATE $\Omega_{A{c_i}}$  $\gets$ AssessTrust$(G', A, c_i, H-1)$
\STATE  $\Omega_i \gets \Delta(\Omega_{A{c_i}},\omega_{{c_i} C})$
\ENDIF
\STATE $n \gets n+1$
\ENDFOR
\IF {$n > 1$}
\STATE $\Omega_{AC} = \Theta (\Omega _1^{} \cdots \Omega _n^{})$
\ELSE
\STATE $\Omega_{AC} = \Omega _n$
\ENDIF
\ELSE
\STATE $\Omega_{AC} =\left<0,0,0\right>$
\ENDIF
\end{algorithmic}
\end{algorithm}
Based on Theorem~\ref{t2}, we design the AssessTrust algorithm, as shown in Algorithm~\ref{A1}. 
The algorithm is based on the 3VSL model and is able to handle any arbitrary network topologies. 
The inputs of AT algorithm include a social network graph $G$, a trustor $A$, a trustee $C$, and the maximum searching depth $H$, measured by number of hops. 
Specifically, $H$ determines the longest distance the algorithm will search between the trustor and trustee. 
$H$ controls the searching depth of the AT algorithm, which is necessary because $G$ could be potentially very large. 

To compute $A$'s individual opinion on $C$, AT applies a recursive depth first search (DFS) on graph $G$, with a maximum searching depth of $H$. 
AT starts from the trustee $C$ and visits all $C$'s incoming neighbors $c_i$'s, as shown in lines $1$ to $12$.
For each node $c_i$, we denote $A$'s opinion on $C$'s trust obtained through $c_i$ as $\Omega_i$.
At this moment, the opinion $\Omega_i$ is unknown unless $c_i$ is the trustor node $A$. 
In this case, we have $\Omega_i = \omega_{{c_i}C} = \Omega_{AC}$. %
Otherwise, the value of $\Omega_i$ needs to be computed recursively by the AT algorithm.
To do so, AT recalls itself on the new graph $G^{\prime}$ that keeps all the edges in the current graph except edge $e(c_i, C)$ and node $C$, as shown in line $7$. 
The output of the AT algorithm, with $G^{\prime}$ as the input graph, will be $A$'s opinion on $c_i$'s trust, as shown in line $9$.
When all the incoming neighbors $c_i$'s are processed, all the edges connecting to $C$ will be removed from the graph as well.
After that, if AT visits $C$ again in the future, \ie, $C$ is involved in a cycle in $G$, the algorithm will stop as there is no incoming neighbor for $C$.
In other words, cycles in graph $G$ will be eliminated when AT searches the graph. 
A cycle involving a node essentially means the node holds a trust opinion about itself, which does not make sense as a node must absolutely trust itself.
Therefore, it is meaningless to let a node to compute its own trust, levering others' opinions upon itself.  

When the input graph becomes $G^{\prime}$, the trustee will be $c_i$ and the maximum searching depth is decreased to $H-1$, as shown in line $8$.
If there are more than one $c_i$, all the resulting opinions $\Omega_i$'s will be combined to yield the opinion $\Omega_{AC}$, as shown in line $14$.
Otherwise, the only obtained opinion $\Omega_{i}$ will be assigned to $\Omega_{AC}$, as shown in line $16$.
In the end, if the searching depth reaches $H$, AT return an empty opinion, as shown in line $19$.

\subsection{Illustration of the AssessTrust Algorithm}
\label{AnEx}

\begin{figure}[!t]
\centering
\subfigure[Bridge topology]{%
\includegraphics[width=1.75in]{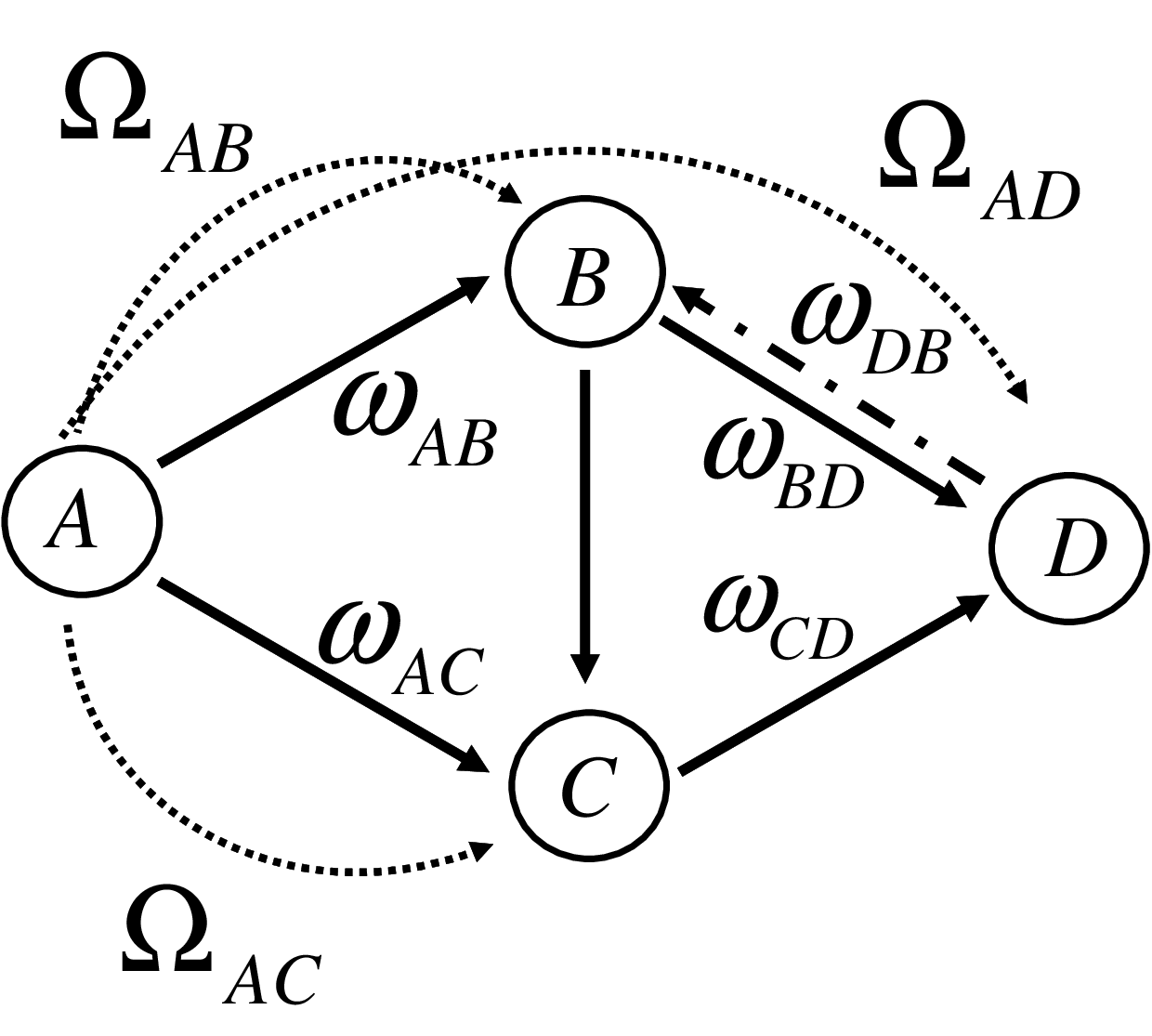}
\label{Bridge_1}
}
\hfil
\subfigure[Decomposition parsing  parsing tree]{%
\includegraphics[width=1.75in]{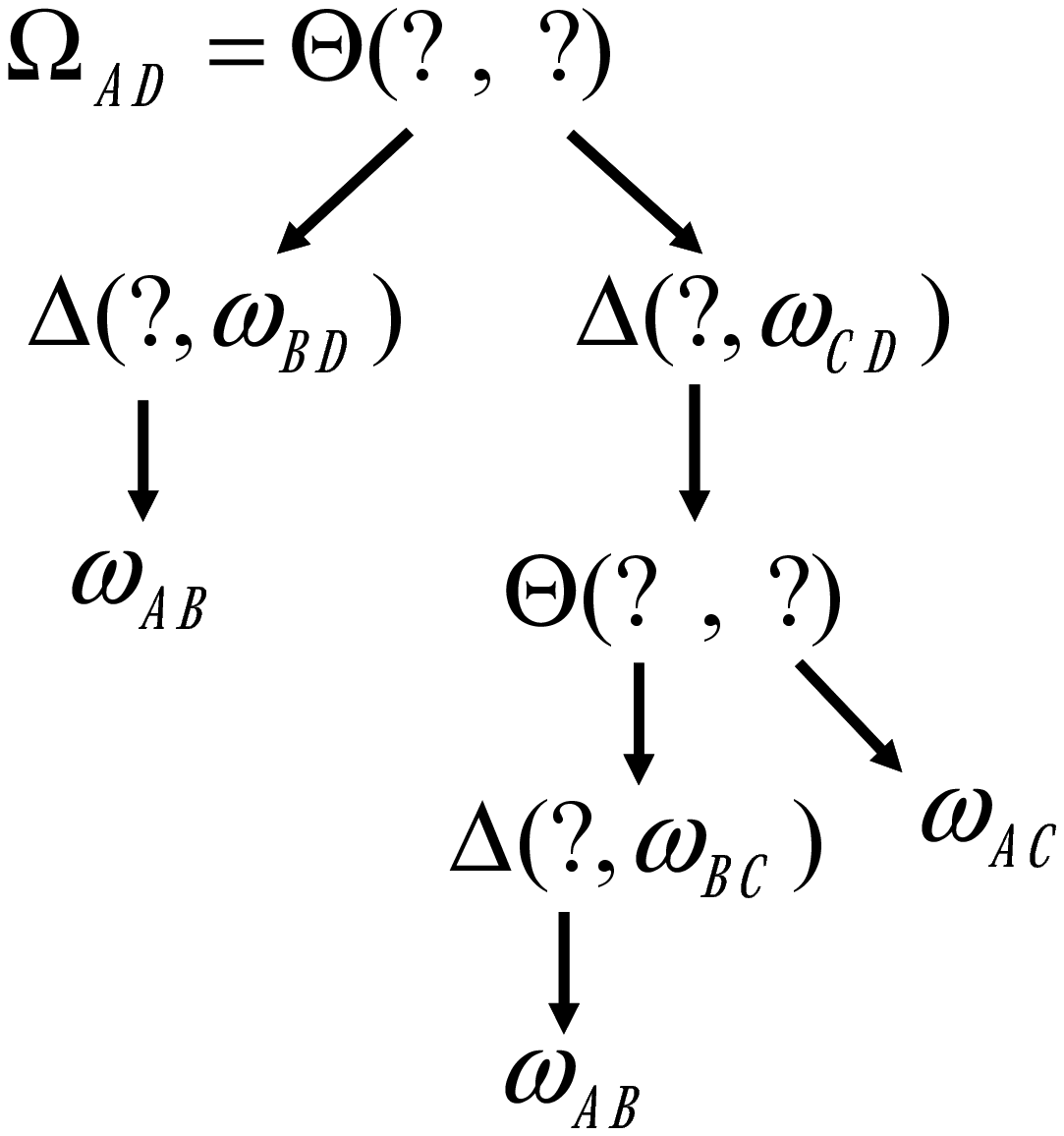}
\label{Bridge_2}
}
\caption{An illustration of 3VSL based on the bridge topology}
\label{Bridge}
\end{figure}

In this section, we will use the bridge topology shown in Fig.~\ref{Bridge_1} to illustrate how the AT algorithm computes $A$'s indirect opinion on $C$, denoted as $\Omega_{AD}$. 
To differentiate from the direct opinion, we use $\Omega$ to denote the indirect opinion.
As shown in Fig.~\ref{Bridge_1}, to compute $\Omega_{AD}$, discounting and combining operations are applied on opinions $\omega_{AB}, \omega_{AD}, \omega_{BD}, \omega_{CD}$, and $\omega_{BC}$. 
AT starts from the trustee $D$ and searches the network backwards, and recursively computes the trust of every node.
As a result, we obtain a parsing tree, shown in Fig.~\ref{Bridge_2}, to indicate the correct order that discounting and combining operations are applied in computing $A$'s opinion on $D$.
By traversing the parsing tree in a bottom-up manner, $A$'s indirect opinion about $D$ can be computed as 
\begin{equation}
\label{eq:atrs}
\Theta \left(\Delta ({\omega _{A{B}}},{\omega _{{B}D}}), \Delta (\Theta (\Delta ({\omega _{A{B}}},{\omega _{{B}{C}}}), {\omega _{AC}}),{\omega _{{C}D}}) \right).
\end{equation}

To understand how exactly AT searches the bridge network, we use $AT^{(k)}(i,j)$ to denote it is for the $k$th time that AT is called, to compute the $i$'s opinion on $j$.
At the first time when AT is called, $A$'s opinion on $D$ is computed from
\[
\Theta \left(\Delta(\Omega_{AB},\omega_{BD}), \Delta(\Omega_{AC}, \omega_{CD}) \right),
\] 
where $\Omega_{AB}$ and $\Omega_{AC}$ are $A$'s indirect opinions on $B$ and $C$, respectively. 
These two opinions will then be computed by $AT^{(2)}(A, B)$ and $AT^{(3)}(A, C)$, respectively.
In $AT^{(3)}(A, C)$, AT computes $A$'s opinion about $C$ as 
\[
\Theta \left(\Delta(\Omega_{AB}, \omega_{BC}), \omega_{AC} \right),
\]
where $\Omega_{AB}$ is computed by $AT^{(4)}(A, B)$.
Finally, $A$'s opinion on $D$ can be computed from Eq.~\ref{eq:atrs}.
In the bridge-topology network, the AT algorithm is called four times in total: $AT^{(1)}(A, D)$, $AT^{(2)}(A, B)$, $AT^{(3)}(A, C)$ and $AT^{(4)}(A, B)$.
Note that the opinion output from $AT(A,B)$ is used twice, \ie, in sub-graphs $A \rightarrow B \rightarrow C$ and $A \rightarrow B \rightarrow D \rightarrow C$, which is allowed in 3VSL.

The AT algorithm still works if a cycle is introduced in the graph, e.g., the edge from $B$ to $D$ is reversed.
With the reversed edge $DB$, a loop $D \rightarrow B \rightarrow C \rightarrow D$ is formed.
In the following, we will show how AT works on the graph with a cycle $D \rightarrow B \rightarrow C \rightarrow D$. 
The algorithm starts from $D$ and visits $C$, and then recalls itself on graph $G^{\prime}$ in which $D$ and edge $CD$ are removed.  
The algorithm then reaches $A$ and $B$.
When it processes $B$, AT cannot visit $D$ as $D$ was already removed, so the algorithm quits. 
As such, the cycle $D \rightarrow B \rightarrow C \rightarrow D$ is eliminated while computing the indirect trust opinion $\Omega_{AD}$.

\subsection{Time Complexity Analysis}
\label{AT:TCA}
In this section, we present the time complexity of the AssessTrust algorithm.
Because AT is a recursive algorithm, the recurrence equation of its time complexity is
\begin{align*}
T(n) &= (n-1) \cdot \left (T(n-1) + {C_1} \right) + {C_2} + O(n-1)\\
 &= (n-1) \cdot T(n-1) + O(n-1) + C,
\end{align*}
where $(n-1)$ is the maximum number of incoming edges to the trustee (line 3), assuming there are $n$ nodes in the network. 
$T(n-1)$ is the time complexity of recursively running AT on each branch (line 8), $C_1$ is the time for lines $4-7$ and $9-11$. $O(n-1)$ is the time for combining operations (line $14$). $C_2$ is the time used outside the ``for'' loop (line $13-20$).
Therefore, the time complexity of AT is
\begin{equation*}
O\left(\sum\limits_{i = 1}^H {\frac{{(n - 1)!}}{{(n - 1 - i)!}}}\right) =  O(n^H),
\end{equation*}
where $H$ is the maximum searching depth, and $n$ is the number of nodes in the network.

\section{Evaluations}
\label{CH:eval}
In this section, we evaluate the properties and performances of the 3VSL model and AT algorithm. 
We conduct comprehensive experiments to evaluate the accuracy of 3VSL model and compare its performance to that of subjective logic, in two real-world datasets: Advogato and PGP.

For the AT algorithm, we evaluate its accuracy and compare its performance to another trust assessment algorithm, called TidalTrust, in Advogato and PGP.
We investigate the reasons why AT outperforms TidalTrust by analyzing the results obtained from these experiments. 

To understand how accurate various models are in assessing trust within OSNs, we adopt F1 score~\cite{f1sc} as the evaluating metric. 
The F1 score is chosen because it is a comprehensive measure for different models in predicting or inferring trust~\cite{f1sc}. 

After evaluating the accuracy of different trust models, we evaluate the performance of the AT algorithm and compare it to these benchmark solutions: TrustRank and EigenTrust. 

\subsection{Dataset}

The first dataset, Advogato, is obtained from an online software development community where an edge from user $A$ to $B$ represents $A$'s trust on $B$, regarding $B$'s ability in software development. 
The trust value between two users is divided into four levels, indicating different trust levels.
The second dataset, Pretty Good Privacy (PGP), is collected from a public key certification network where an edge from user $A$ to $B$ indicates that $A$ issues a certificate to $B$, \ie, $A$ trusts $B$.
Similar to Advogato, the trust value is also divided into four levels.

According to the document provided by Advogato, a user determines the trust level of another user, based on only certain evidence.  
Therefore, a low-trust edge in Advogato indicates an opinion that contains negative evidence. 
On the other hand, in PGP, a user tends to give a low trust certification if he is not sure whether the other user is trustworthy or not. 
A user in PGP will never give a certification to anyone who has malicious behavior. 
Therefore, a low trust level in PGP indicates an opinion that contains uncertain evidence. 
We select these two datasets because they are obtained from real world OSNs where trust relations between users are quantified as non-binary values. 
In addition, the different definitions of trust in these two datasets allow us to evaluate the performance of 3VSL in different trust social networks.    
Statistics of these datasets are summarized in Table~\ref{tab:stat}.

\begin{table}[h]
\caption[Statistics of the Advogato and PGP datasets]{Statistics of the Advogato and PGP datasets.}
\label{tab:stat}
\centering
\begin{tabular}{|c|c|c|c|c|}
\hline
\bf Dataset   & \bfseries \# Vertices  & \bfseries  \# Edges  & \bfseries Avg Deg  & \bfseries Diameter   \\
\hline
Advogato &  6,541 &   51,127  &  19.2 &  4.82  \\
\hline
PGP & 38,546 & 31,7979 &  16.5 &  7.7 \\
\hline
\end{tabular}
\end{table}

\vspace{-7mm}
\subsection{Dataset Preparation}
In Advogato, trust is classified into four ordinal levels: \textit{observer}, \textit{apprentice}, \textit{journeyer} and \textit{master}. 
Similarly, in PGP,  trust is classified into four levels: \textit{0}, \textit{1}, \textit{2} and \textit{3}. 
Both Advogato and PGP provide directed graphs where users are nodes and edges are the trust relations among users.  
Because the trust levels are in ordinal scales, a transformation is needed to convert a trust level into a trust value, ranging from $0$ to $1$.

In the experiments, we set the total evidence values $\lambda$ as $10$, $20$, $30$, $40$, and $50$.
Given a certain $\lambda$, we can represent an opinion as $\left<  \frac{\alpha }{{\lambda }}, \frac{\beta }{{\lambda  }}, \frac{\gamma }{{\lambda }} \right>$. 
As aforementioned, the meanings of trust in Advogato and PGP are different, so we use different methods to construct opinions in Advogato and PGP. 
We assume the opinions in Advogato only contain positive and negative evidence, \ie, $\gamma = 0$.
Therefore, an opinion of 3VSL in Advogato can be expressed as $\left< \alpha,\lambda \left(1 - \frac{\alpha }{{\lambda }} \right),0 \right>$.
Given the total number of evidence value $\lambda$, an opinion in Advogato is in fact determined by $\frac{\alpha }{{\lambda }}$, \ie, the proportion of positive evidence.
To properly set the value of $ \frac{\alpha }{{\lambda }}$, we use the normal score transformation technique~\cite{powers2008statistical} to convert ordinal trust values into real numbers, ranging from 0 to 1.  
Specifically, trust levels are first converted into z-scores by the normal score transformation method, based on their distributions in the datasets. 
Then, we map the z-scores to different $\frac{\alpha }{{\lambda }}$'s, according to the differences among the z-scores.
For example, the \textit{master} level trust is converted into $( \frac{\alpha }{{\lambda }})_{3} = 0.9$. 
For the \textit{observer} level trust, we use different values of $(\frac{\alpha }{{\lambda }})_{0}$ as $0.1$, $0.2$, $0.3$, $0.4$ and $0.5$ to indicate the possible lowest trust levels.  
With the highest and lowest values of  $\frac{\alpha }{{\lambda }}$, we interpolate the values of $( \frac{\alpha }{{\lambda }})_{1}$ and $(\frac{\alpha }{{\lambda }})_{2}$ for \textit{apprentice} and \textit{journeyer} level trusts, based on the intervals between the corresponding z-scores.
Because there are five different $\lambda$'s and five different $( \frac{\alpha }{{\lambda }})_{0}$'s, we have a total of $25$ combinations of parameters.

For the PGP dataset, we assume there is only positive and uncertain evidence, so we set $\beta = 0$.
Therefore, an opinion of 3VSL in PGP can be expressed as $\left< \alpha ,0, \lambda(1 - \frac{\alpha }{{\lambda }})\right >$.
Similar to Advogato, an opinion in PGP is determined by $\lambda$ and  $ \frac{\alpha }{{\lambda }}$. 
We use the same transformation method to convert the trust relations in PGP into opinions.

\subsection{Accuracy of 3VSL Model}

\begin{figure}
\centering
\subfigure[Advogato]{\includegraphics[width=3in]{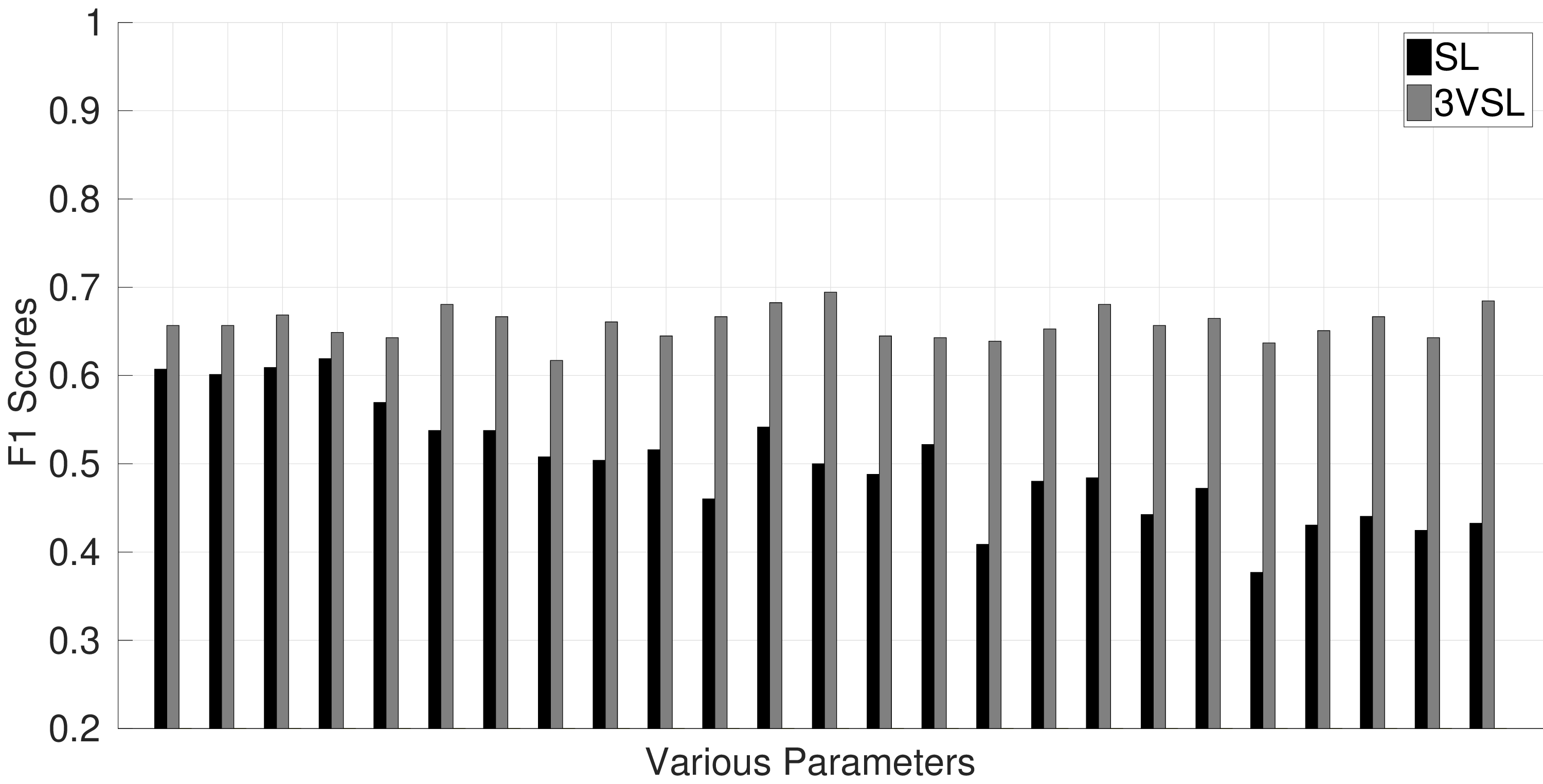}
\label{advg-f1-slcomp}}
\subfigure[PGP]{
\includegraphics[width=3in]{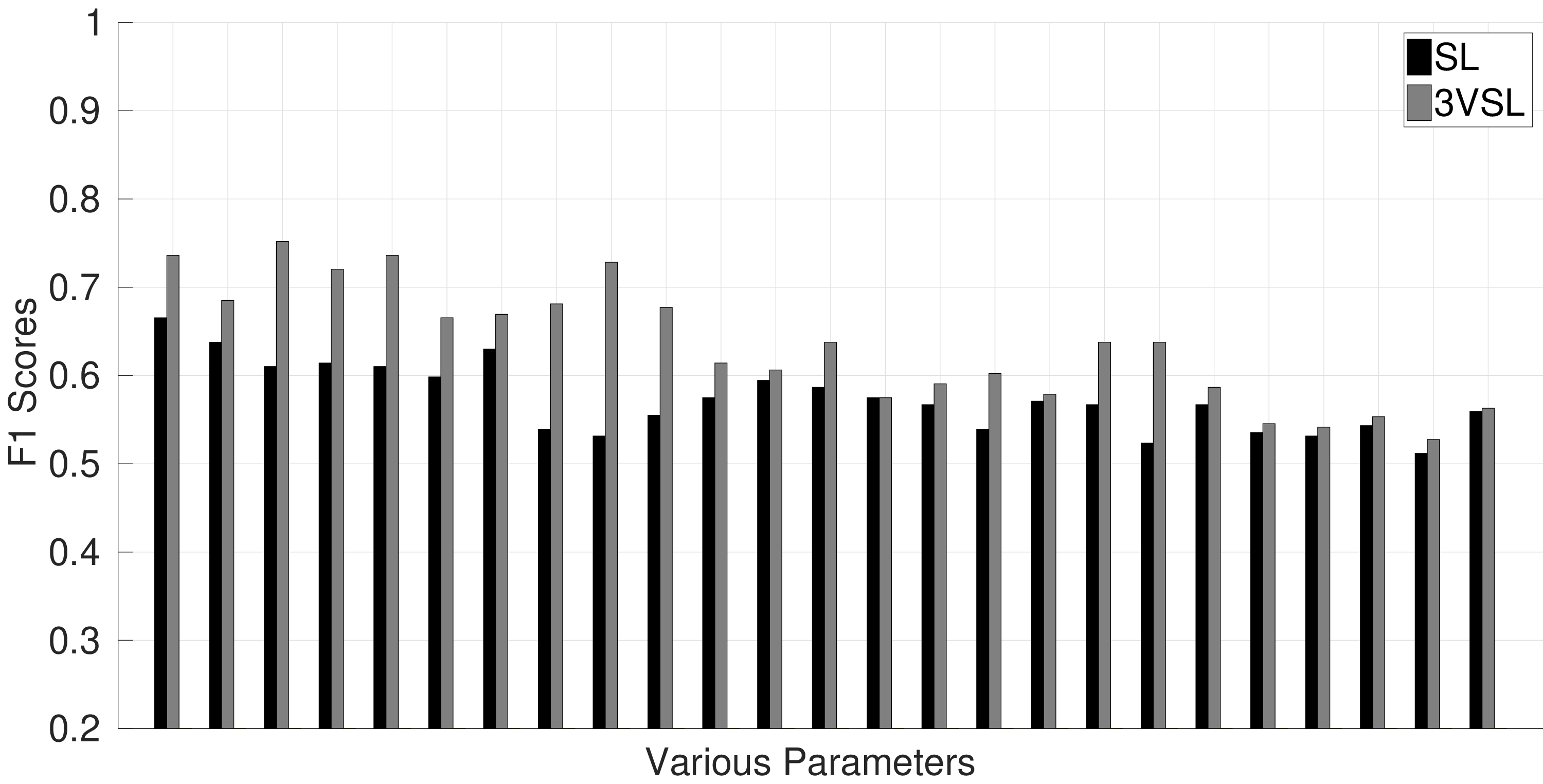}
\label{pgp-f1-slcomp}}
\caption[Histogram of the errors generated by TT, SL* and AT using the Advogato dataset]{F1 scores of 3VSL and SL using the A) Advogato and B) PGP dataset. Parameters are the combinations between base trust levels ($0.1$, $0.2$, $0.3$, $0.4$ and $0.5$) and total evidence values ($10$, $20$, $30$, $40$, and $50$)}
\vspace{-0.2in}
\end{figure}

With the above-mentioned two datasets, we evaluate the accuracy of the 3VSL model.
We also compare the accuracy of the 3VSL model to the SL model.
As we know, SL does not model the trust propagation process correctly and its performance will degrade drastically in real-world OSNs.  
Due to this issue, SL cannot handle social networks with complex network topologies.
Although some approximation solutions are proposed, \eg, removing edges in a social network to reduce it into a simplified graph, there is no existing algorithm that implements any of these solutions.
To make a fair comparison, we design an algorithm called SL*, based on the AT algorithm.
The structure of the SL* algorithm is exactly the same as AT's, however, the discounting and combining operations used in the AT algorithm are replaced with those defined in SL.
As such, SL* implements the SL model and is able to work on OSNs with arbitrary topologies.

The experiments are conducted as follows. 
First, we randomly select a trustor $u$ from the datasets and find one of its $1$-hop neighbors $v$. 
We take the opinion from $u$ to $v$ as the ground truth, \ie., how $u$ trusts $v$. 
Then, we remove the edge $(u,v)$ from the datasets, if there is a path from $u$ to $v$.
%
%
We run the above-mentioned algorithms to compute $u$'s opinion of $v$'s trustworthiness. 
Finally, we compare the computed results to the ground truth. 
We select $200$ pairs of $u$ and $v$ to get statistically significant results. 
To compare the computed results to the ground truth, we first use the expected beliefs of computed opinions as the trust values in 3VSL and SL.
Then, we round the expected beliefs to the closest trust levels based on the ground truths. 
Finally, we use F1 score to evaluate the accuracy of different models. 
Because we do not know the correct parameter settings, we test the above-mentioned $25$ combinations of parameters to conduct a comprehensive evaluation. 
As shown in Fig.~\ref{advg-f1-slcomp} and~\ref{pgp-f1-slcomp}, 3VSL achieves higher F1 scores than SL, with all different parameter settings, in both datasets.   
Specifically, 3VSL achieves F1 scores ranging from $0.6$ to $0.7$ in Advogato, and $0.55$ to $0.75$ in PGP. 
On the other hand, the F1 scores of SL range from $0.35$ to $0.6$ in Advogato and $0.55$ to $0.67$ in PGP. 
Considering F1 score is within the range of $[0,1]$, we conclude that 3VSL significantly outperforms SL.

More importantly, we observe that the F1 scores of 3VSL are relatively stable,  with different parameter settings.
However, the F1 scores of SL fluctuate, indicating SL is significantly affected by the parameter settings.
Overall, we conclude that 3VSL is not only more accurate than SL but also more robust to different parameter settings.

\begin{figure}[h]
\vspace{-0.1in}
\centering
\includegraphics[width=3in]{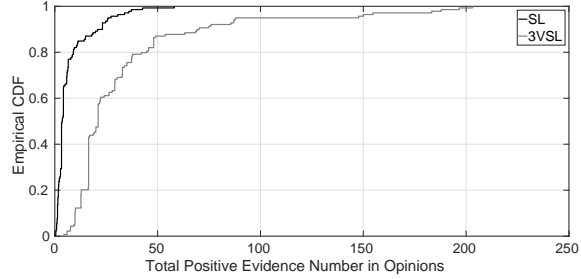}
\caption[CDFs of $\alpha + \beta$ in opinions computed by 3VSL and subjective logic using the  Advogato dataset]{CDFs of $\alpha + \beta$ in opinions computed by 3VSL and subjective logic using the  Advogato dataset.}
\vspace{-0.1in}
\label{advg_evid_num}
\end{figure}

We further investigate the reason why 3VSL outperforms SL by looking at the evidence values in the resulting opinions, computed by 3VSL and SL.
We choose the results from experiments with the parameter setting (0.3, 30), wherein 3VSL performs the best.
We are only interested in the cases where 3VSL obtains more accurate results than SL.  
We measure the values of certain evidence ($\alpha + \beta$) in the resulting opinions computed by 3VSL and SL.
The CDFs of the values of certain evidence are then plotted in Fig.~\ref{advg_evid_num}.
As shown in Fig.~\ref{advg_evid_num}, the values of  $(\alpha + \beta)$ in the opinions computed by SL are much lower than that of 3VSL.
It results in a lack of evidence in computing the expected beliefs of opinions by SL.  
This observation matches the example introduced in Section~\ref{advtg_3vsl}.
Because 3VSL employs a third state to store the uncertainty generated in trust propagation, it is more accurate in modeling and computing trust in OSNs. 
\subsection{Performance of the AssessTrust Algorithm}

\begin{table}[t!]
\centering
\begin{tabular}{ |p{1cm}|p{2.6cm}|p{2.6cm}|p{3cm}|p{2.6cm}|  }
\hline
    & Advogato & PGP \\
\hline
 AT & $(0.3,30)$ & $(0.1,30)$ \\
 \hline
 SL*  & $(0.3,30)$ & $(0.1,30)$  \\
\hline
 TT  & $(0.2, -)$ & $(0.1,-)$ \\
\hline
\end{tabular}
\caption[Selected parameters (base trust level, total evidence value) for AT, SL* and TT]{Selected parameters (base trust level, total evidence value) for AT, SL* and TT. Note that TT employs a number to represent trust, so its evidence value is empty.}
\label{advg:pmt}
\end{table}

\begin{figure}
\vspace{-0.15in}
\centering
\subfigure[Advogato]{\includegraphics[width=3in]{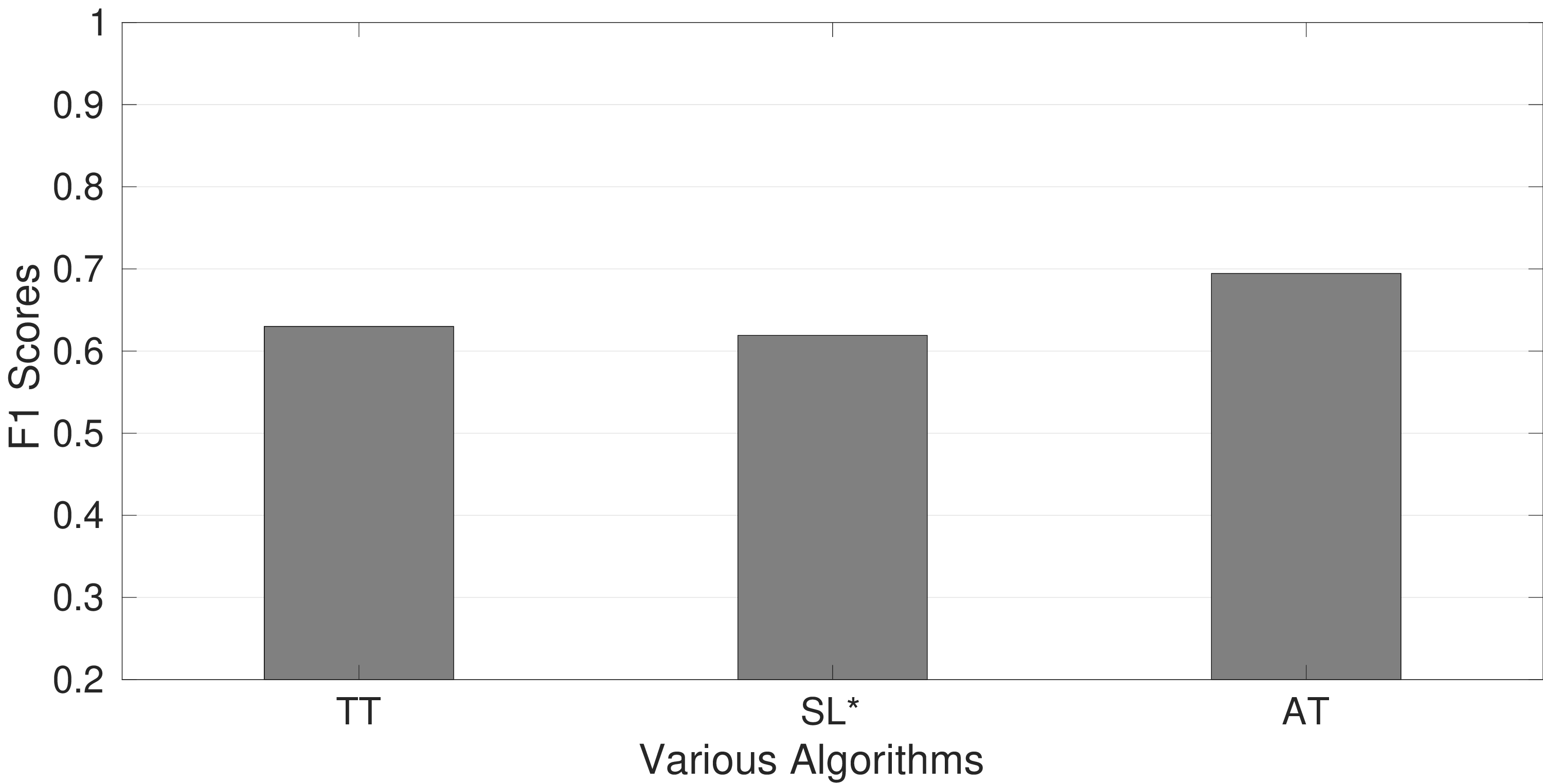}
\label{f1_comp_tt_advg}}
\subfigure[PGP]{
\includegraphics[width=3in]{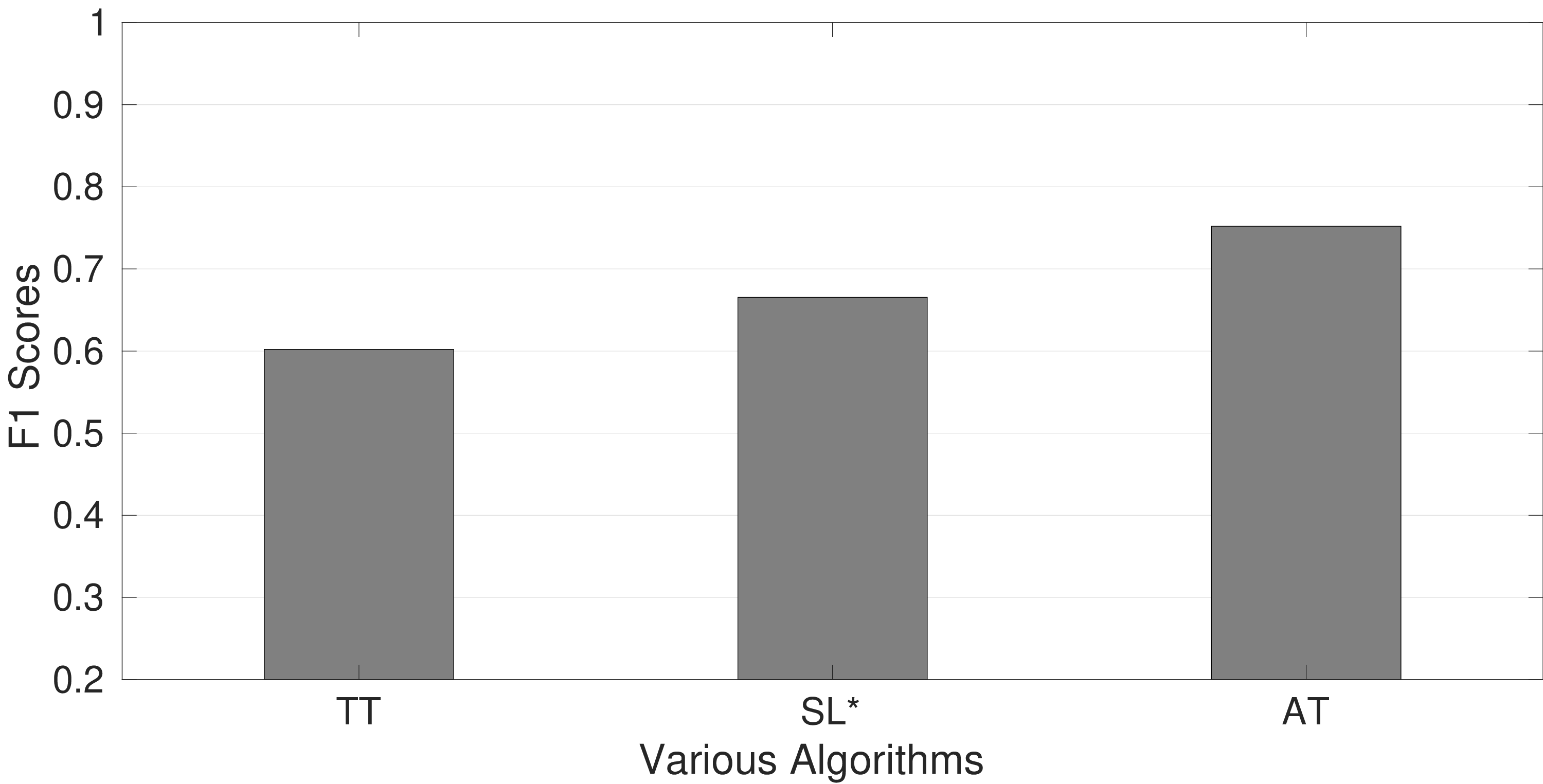}
\label{f1_comp_tt_pgp}}
\caption[Histogram of the errors generated by TT, SL* and AT using the Advogato dataset]{F1 scores of the trust assessment results generated by TT, SL* and AT using the  a) Advogato and b) PGP datasets.}
\vspace{-0.2in}
\end{figure}

After validating the 3VSL model, we study the performance of the AT algorithm and compare it to other benchmark algorithms, including TidalTrust (TT)~\cite{Golbeck:2005:CAT:1104446}, TrustRank (TR)~\cite{Gyongyi:2004:CWS:1316689.1316740} and EigenTrust (ET)~\cite{Kamvar:2003:EAR:775152.775242}.
TidalTrust is designed to compute the absolute trust of any user in an OSN.
However, TR and ET are used to rank users in an OSN based on their relative trustworthiness, \ie, it does not compute the absolute trust.

\begin{figure}
\centering
\subfigure[Error histogram of TT using the Advogato dataset]{\includegraphics[width=3in]{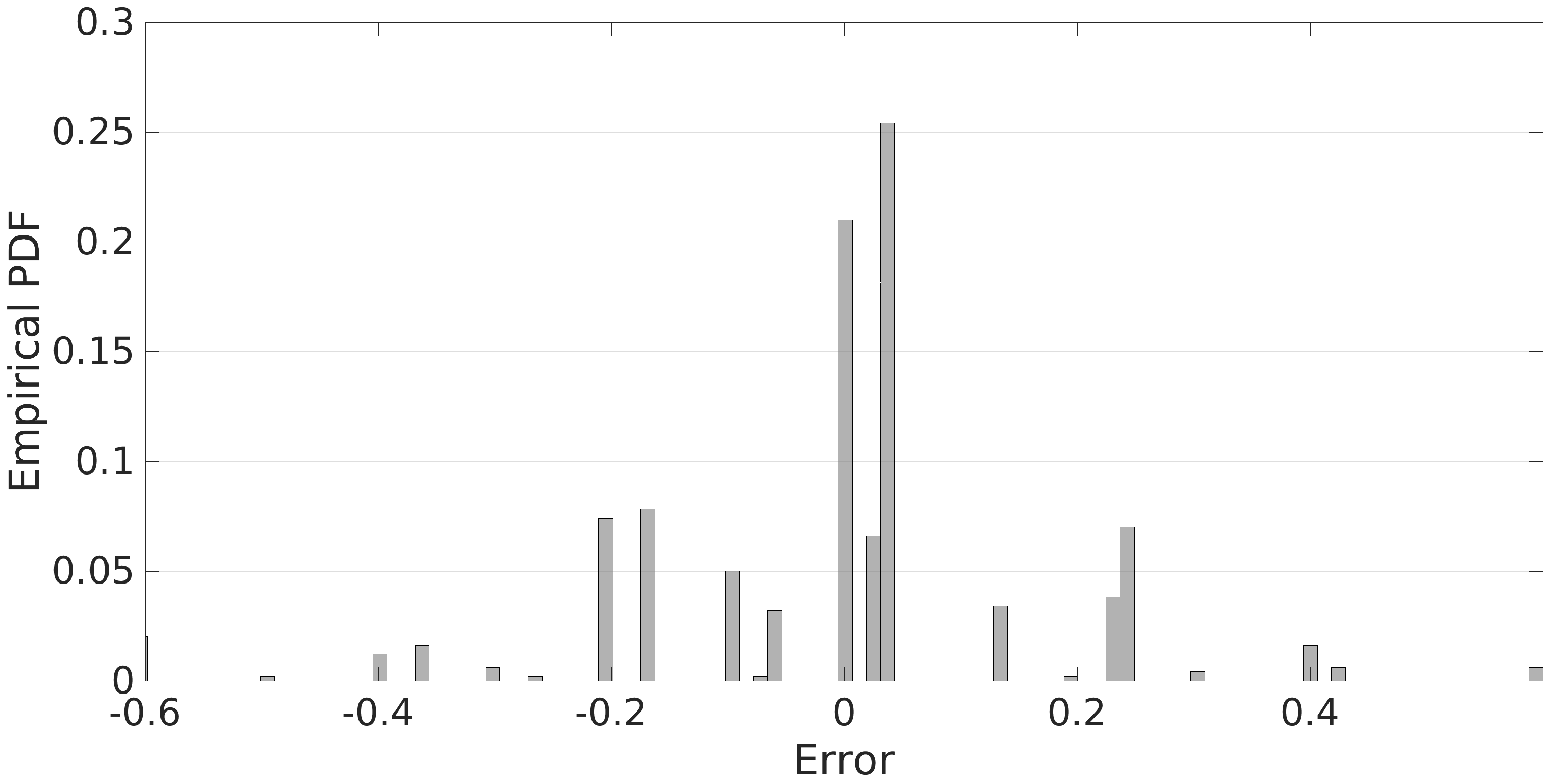}
\label{tt-advg-err}}
\subfigure[Error histogram of TT using the PGP dataset]{\includegraphics[width=3in]{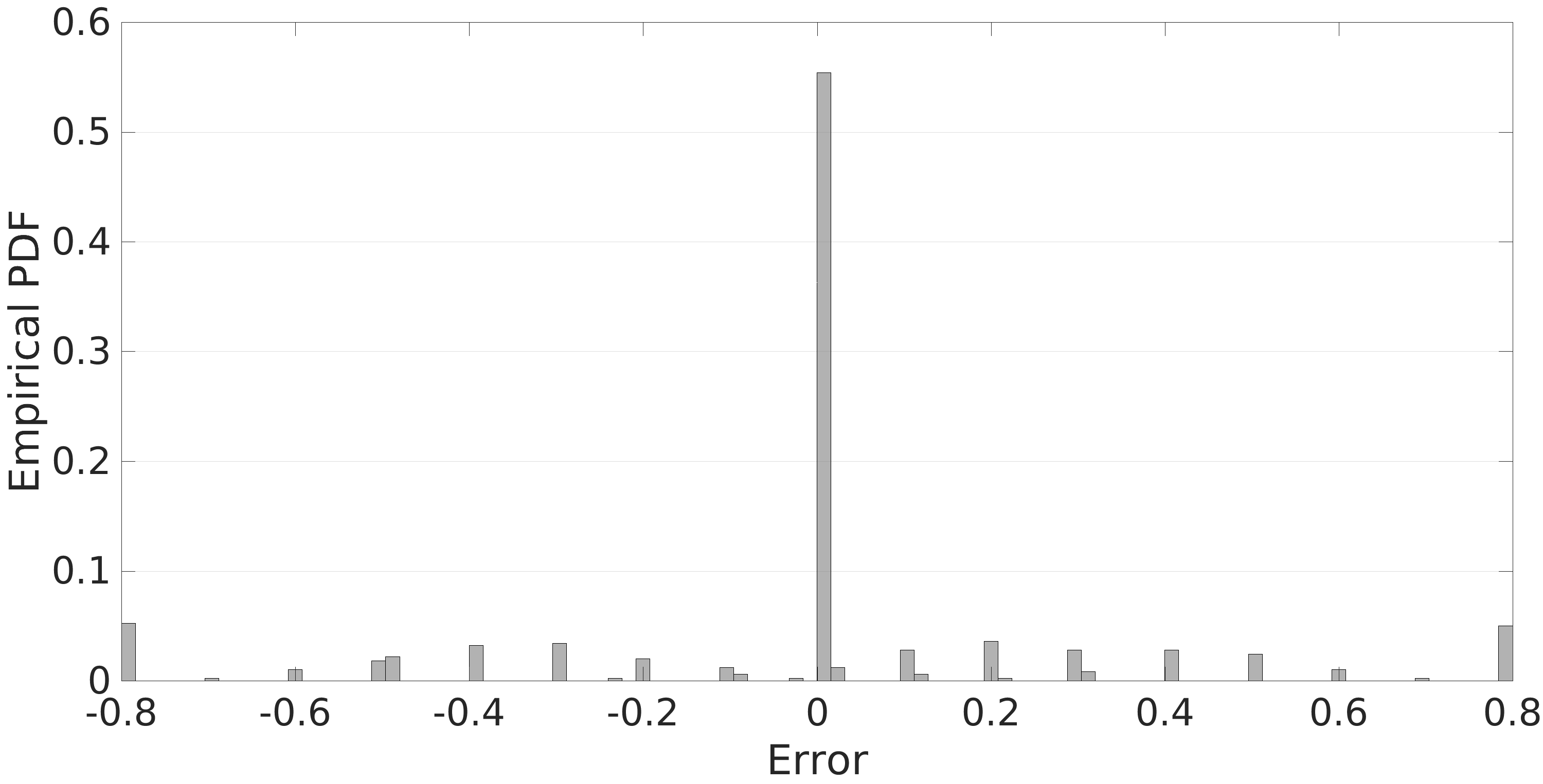}
\label{tt-pgp-err}}
\subfigure[Error histogram of SL* using the Advogato dataset]{
\includegraphics[width=3in]{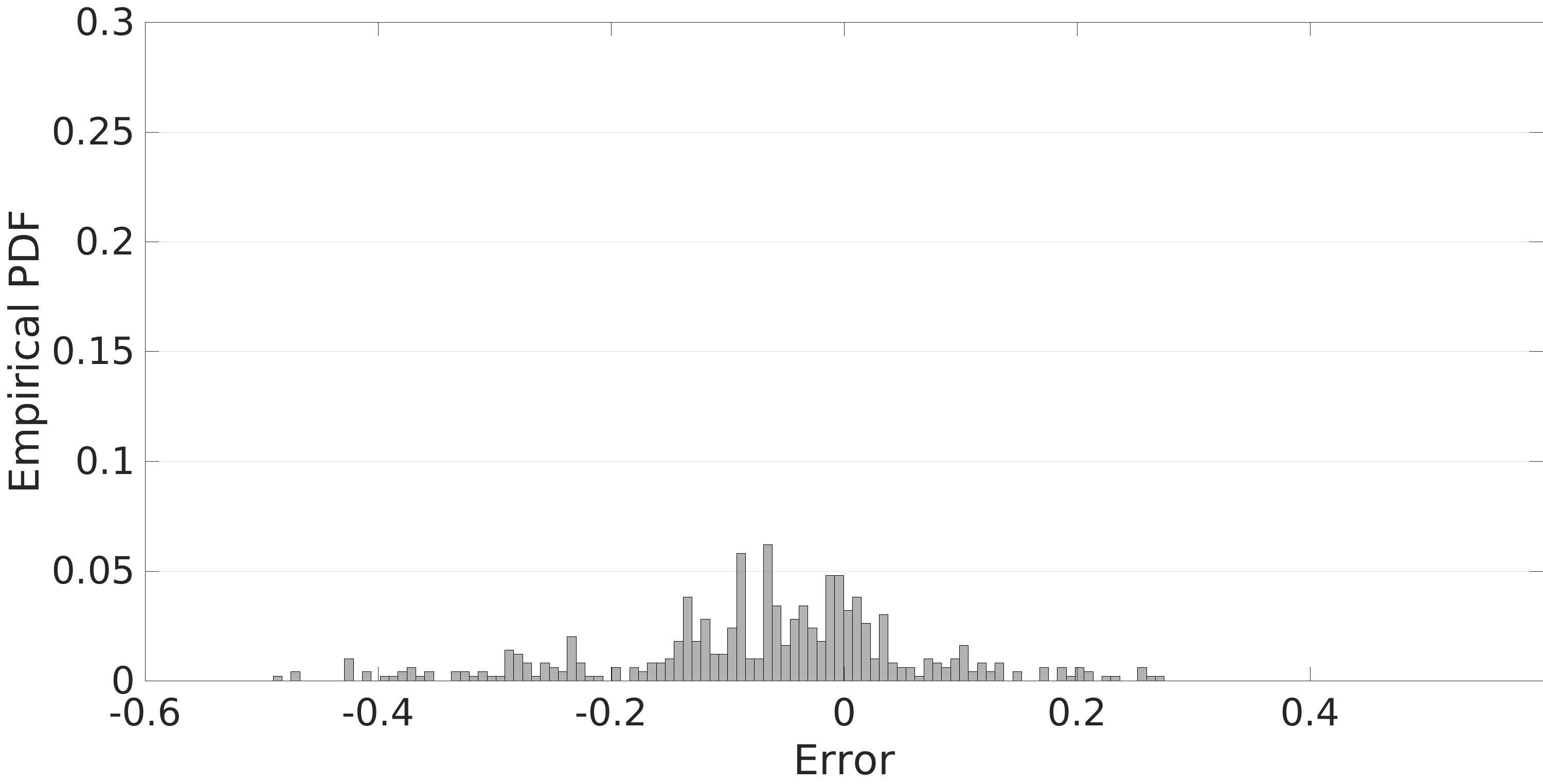}
\label{sl-advg-err}}
\subfigure[Error histogram of SL* using the PGP dataset]{
\includegraphics[width=3in]{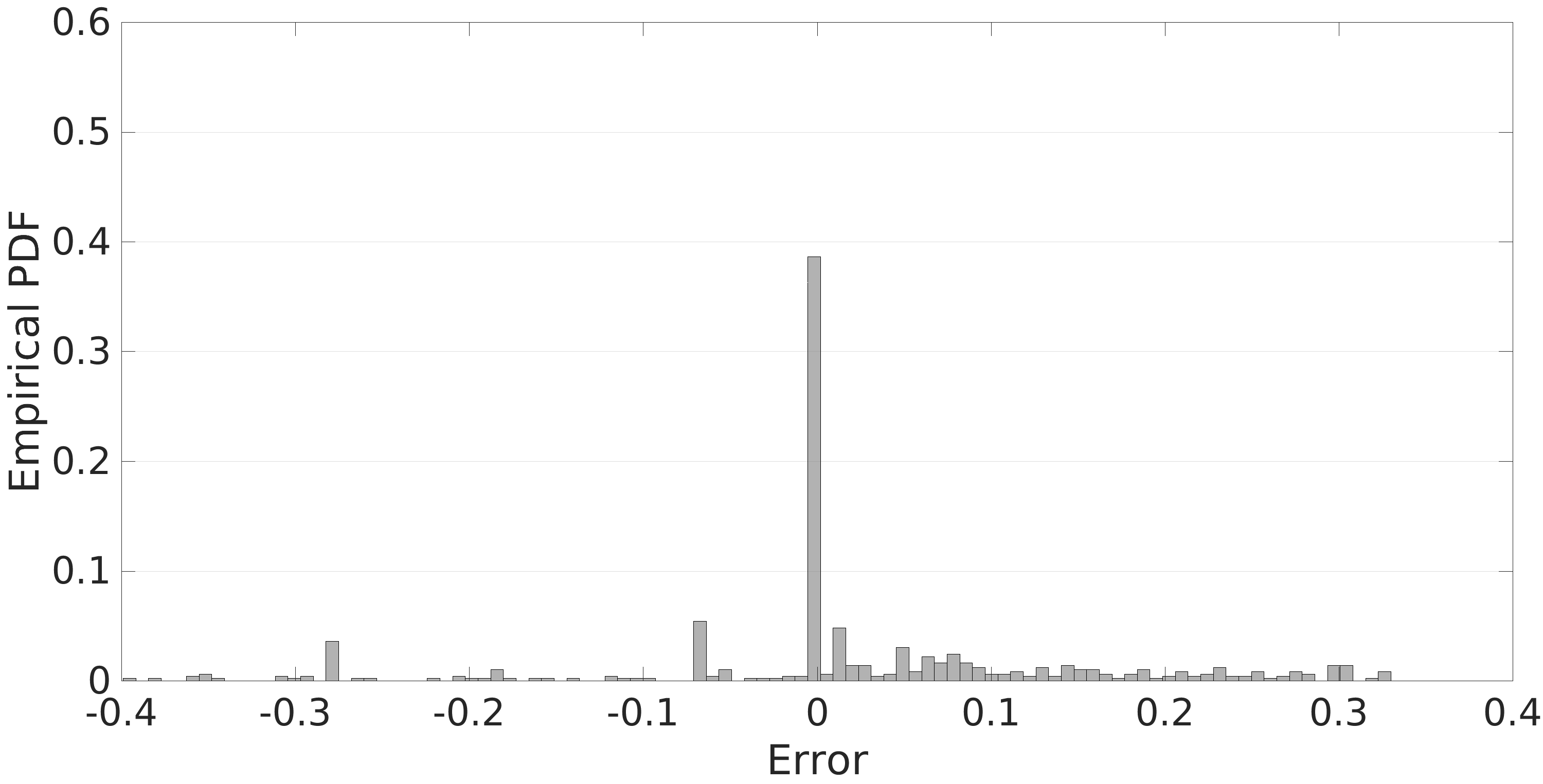}
\label{sl-pgp-err}}
\subfigure[Error histogram of AT using the Advogato dataset]{
\includegraphics[width=3in]{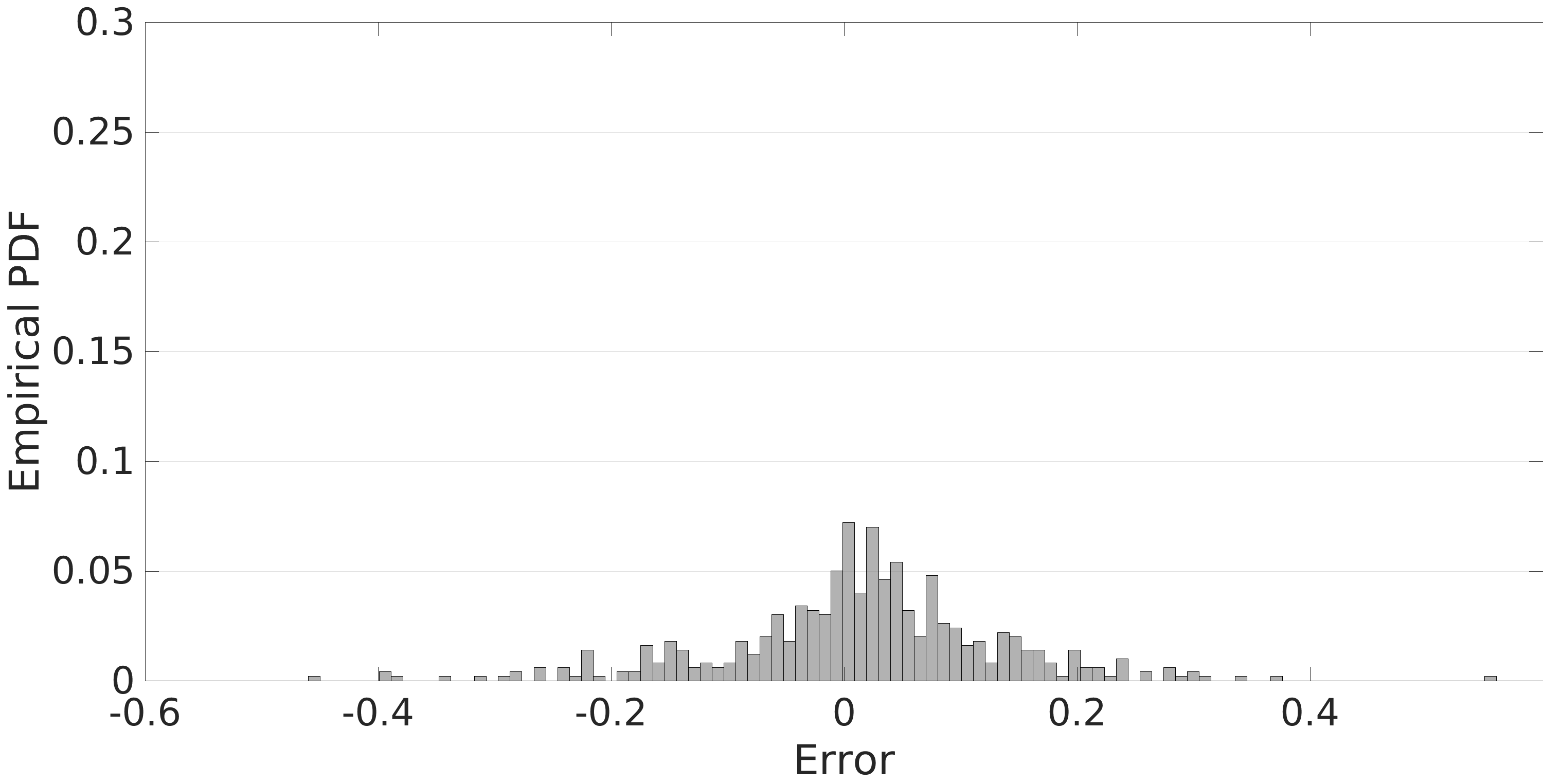}
\label{3vsl-advg-err}}
\subfigure[Error histogram of AT using the PGP dataset]{
\includegraphics[width=3in]{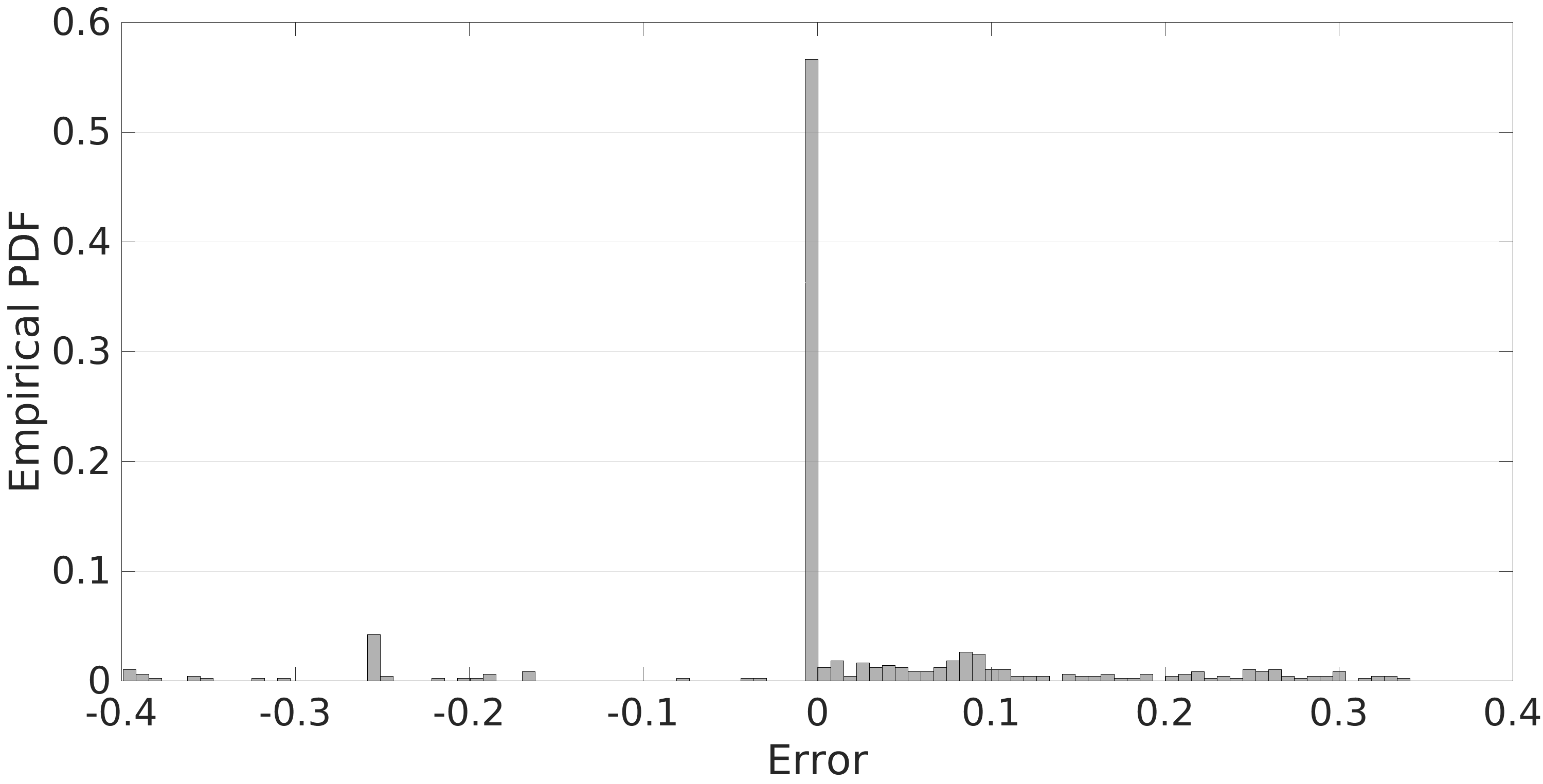}
\label{3vsl-pgp-err}}
\caption[Histogram of the errors generated by TT, SL* and AT using the Advogato dataset]{Histogram of the errors generated by TT, SL* and AT using the Advogato and PGP dataset.}
\label{err_advg_comp_tt}
\vspace{-0.1in}
\end{figure}

Because different benchmark algorithms solve the trust assessment problem differently, we conduct two groups of experiments. 
In the first group of experiments, we compare the performance of AT, SL* and TT in computing the absolute trustworthiness of users in an OSN.
In the experiments, we randomly select a trustor $u$ from the datasets and choose one of its $1$-hop neighbors $v$. 
We take the opinion from $u$ to $v$ as the ground truth. 
Then, we remove the edge $(u,v)$ from the datasets, if there exist paths from $u$ to $v$ in the network.
We run the AT, SL* and TT algorithms to compute the trustworthiness of $v$, from $u$'s perspective. 
Finally, we compare the computed trustworthiness to the ground truth.  

Different parameters will affect the performances of various algorithms, so we choose different parameters for AT and TT so that they can perform well in the experiments.
Because we already validated that 3VSL outperforms SL, regardless of the parameter settings, we choose the same parameter setting used by AT for SL*.
The parameter settings for different algorithms in different datasets are shown in Table~\ref{advg:pmt}.

\begin{figure}
\centering
\subfigure[Advogato]{\includegraphics[width=3in]{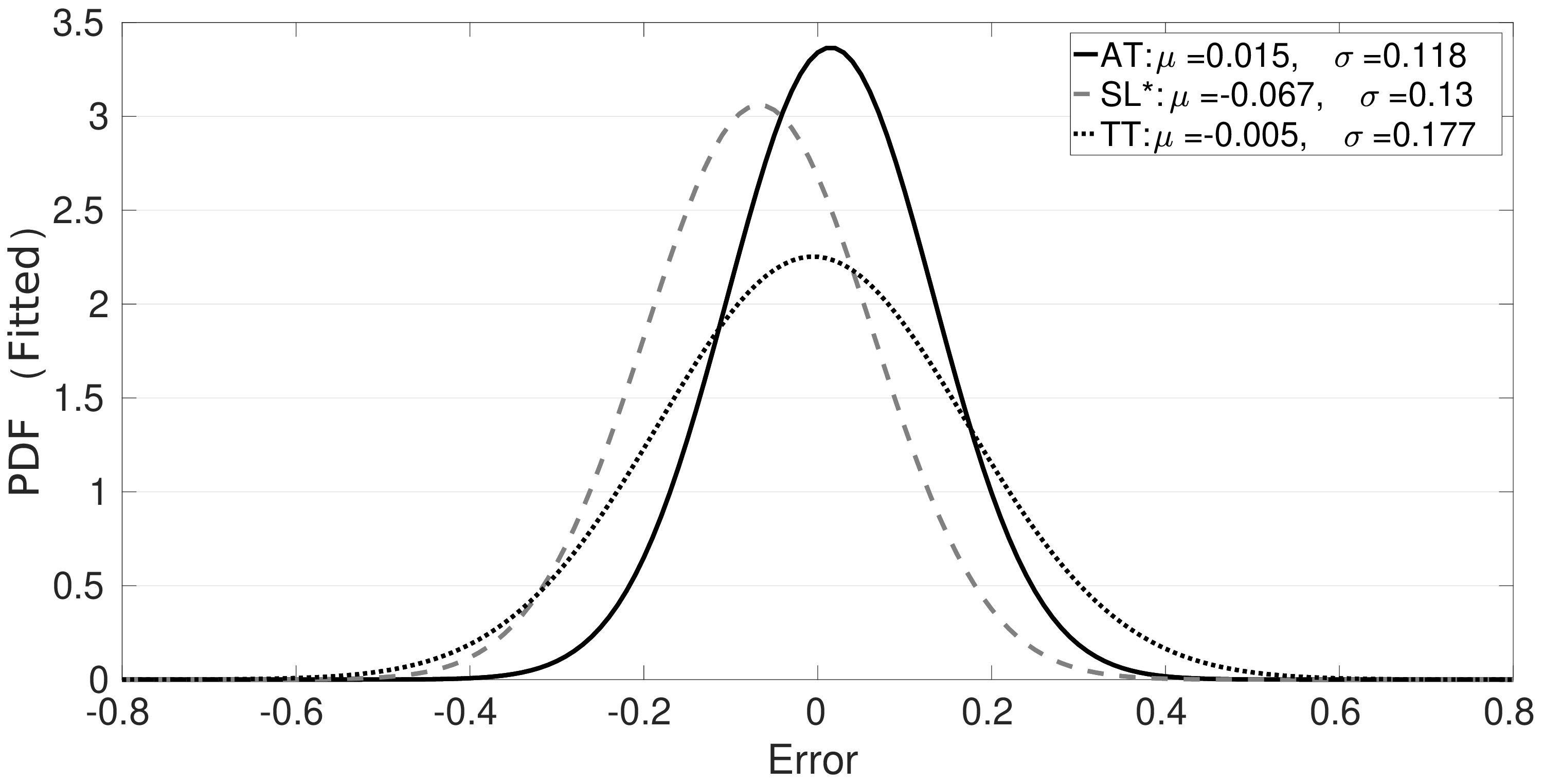}
\label{advg-fit}}
\subfigure[PGP]{
\includegraphics[width=3in]{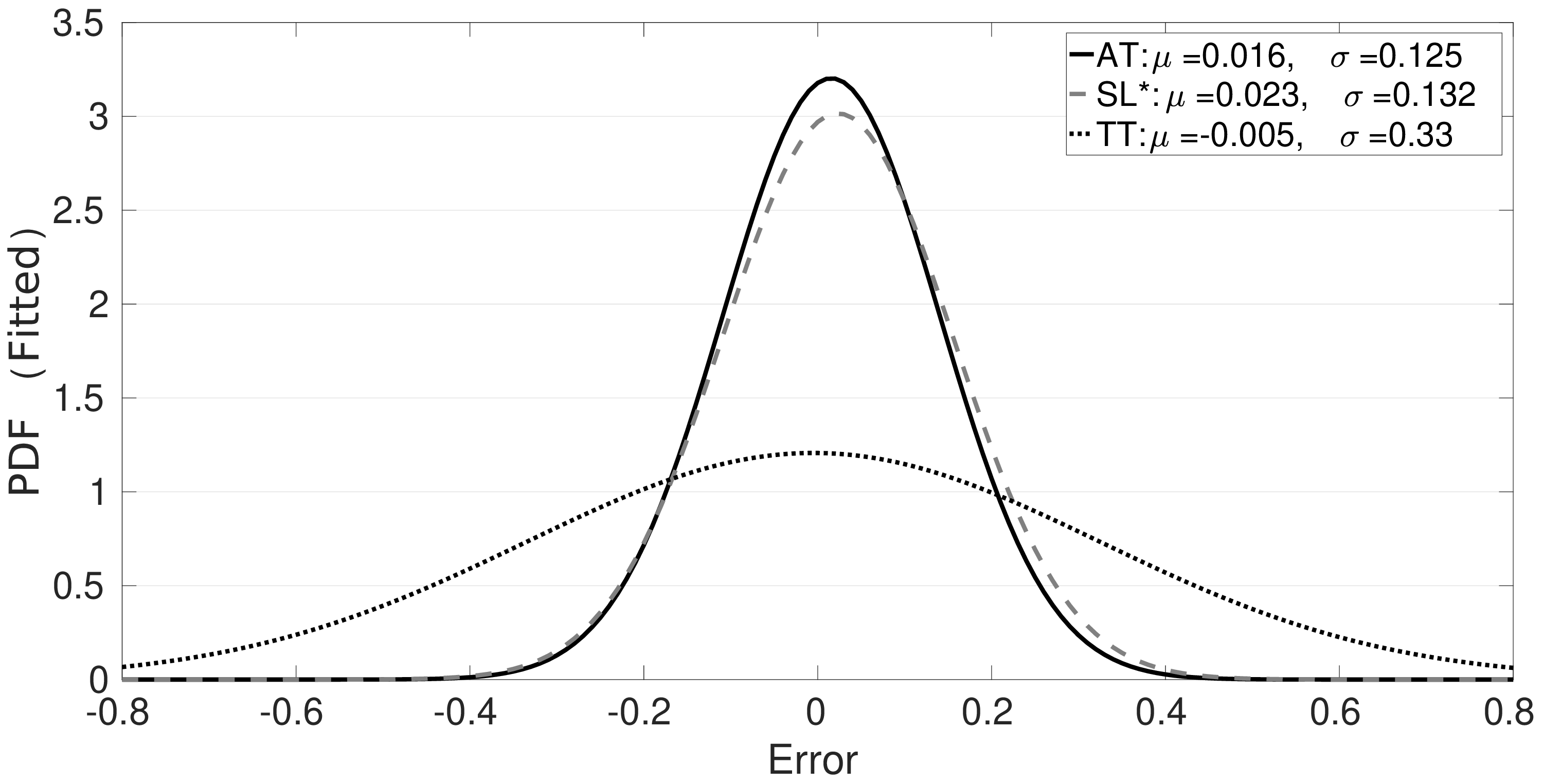}
\label{pgp-fit}}
\vspace{-0.1in}
\caption[Histogram of the errors generated by TT, SL* and AT using the Advogato dataset]{Fitted curves of the error distributions of TT, SL* and AT using the a) Advogato and b) PGP dataset.}
\vspace{-0.2in}
\end{figure}

We first look at the F1 scores of the trust assessment results generated by the three algorithms.
The F1 scores are plotted in Figs.~\ref{f1_comp_tt_advg} and~\ref{f1_comp_tt_pgp}.
As shown in Figs.~\ref{f1_comp_tt_advg} and~\ref{f1_comp_tt_pgp}, AT outperforms TT in both datasets, \ie, TT achieves $0.617$ and $0.605$ F1 scores, and AT offers $0.7$ and $0.75$ F1 scores in Advogato and PGP. 
It is worth mentioning that SL* gives the worst F1 scores, indicating that the problem of subjective logic in modeling uncertainty seriously impacts its performance.  
Besides F1 scores, we also study the distribution of errors in trust assessment results.
The error here is defined as the difference between the computed trust value and the ground truth.  
The error distributions of different algorithms are shown in Figs.~\ref{err_advg_comp_tt}.

From Fig.~\ref{tt-advg-err}, we can see that the errors of TT algorithm is either very small or very large when it is used to assess trust using the Advogato dataset. 
For the SL* and AT algorithms, however, the errors are more concentrated around 0, as shown in Figs.~\ref{sl-advg-err} and~\ref{3vsl-advg-err}.
If the PGP dataset is used, we observe the same phenomena, as shown in Figs.~\ref{tt-pgp-err},~\ref{sl-pgp-err} and~\ref{3vsl-pgp-err}.

\begin{figure}
\centering
\subfigure[Advogato]{\includegraphics[width=3in]{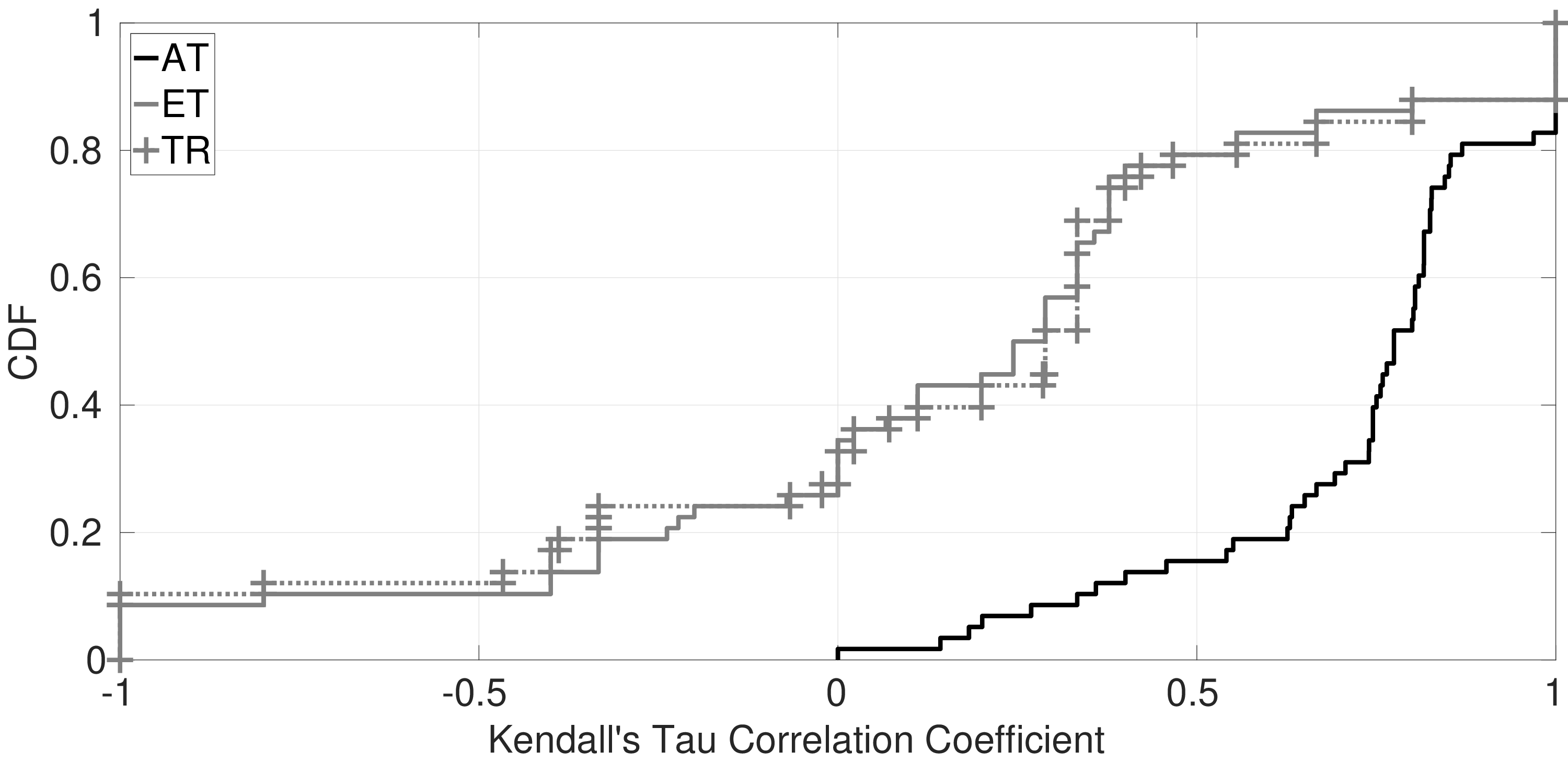}
\label{fig:ranking_advg}}
\subfigure[PGP]{
\includegraphics[width=3in]{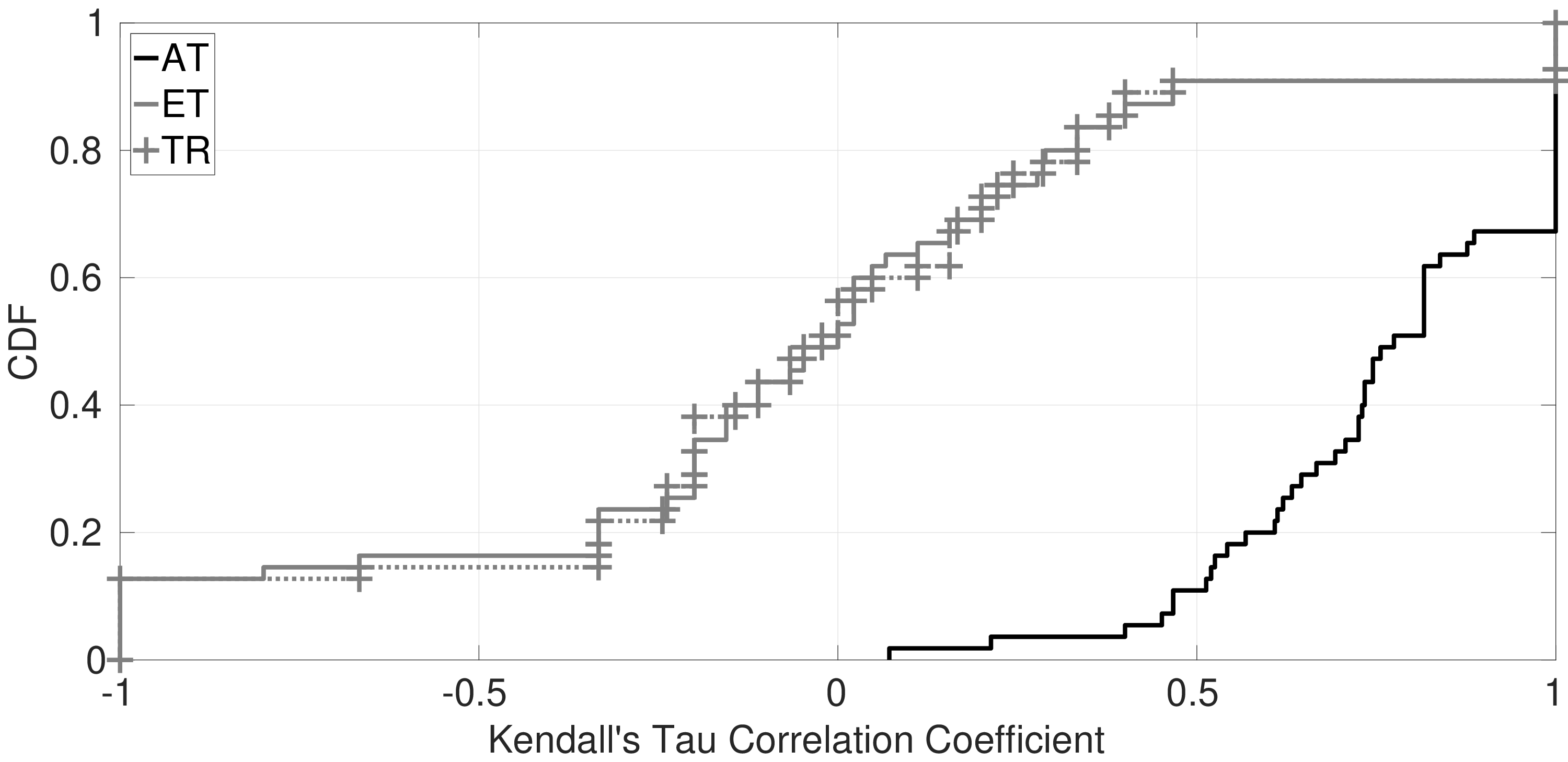}
\label{fig:ranking_pgp}}
\vspace{-0.1in}
\caption[Histogram of the errors generated by TT, SL* and AT using the Advogato dataset]{The CDFs of Kendall's tau ranking correlation coefficients of different algorithms using the a) Advogato and b) PGP dataset.}
\vspace{-0.2in}
\end{figure}

We further fit this histogram data using the Normal Distribution.
As shown in Figs~\ref{advg-fit} and~\ref{pgp-fit}, the fitted curves of the error distributions of different algorithms clearly indicate that AT gives the best trust assessment results.
In these figures, we can see that the error distribution of TT has a close-to-zero mean, \ie, $0.005$ for both datasets, but a large variance.     
On the contrary, the fitted curves of the error distributions of SL* show that SL* has a smaller variance but a large mean, \ie, $0.067$ in Advogato and $0.016$ in PGP.
The fitted curves of the error distributions of AT give the best results, \ie, with a mean of $0.015$ in Advogato and $0.016$ in PGP, and a smaller variance in both datasets. 

In the second group of experiments, we evaluate the performance of AT, ET and TR, in terms of ranking users based on their trustworthiness.
We first randomly select a seed node $u$, and find all its 1-hop neighbors, denoted as $V$. 
Then, we rank the nodes in $V$ based on $u$'s direct opinions on these nodes, \ie, nodes with higher trust values are ranked in higher positions than those with lower trust values.
We take this ranking as the ground truth. 

For each node $v \in V$, we remove edge $(u,v)$ from the datasets if there exist paths from $u$ to $v$.
We run the AT, ET and TR algorithms to compute the trustworthiness of node $v$, from the perspective of $u$. 
Then, we rank the nodes in $V$ based on the expected beliefs of $\omega_{uv}$'s for all possible $v$'s.
We compare the ranking results obtained by the three algorithms to the ground truth.
Here, ranking errors are measured by Kendall's tau ranking correlation coefficients between the computed ranking results and the ground truth. 
We repeat each experiment $100$ times in Advogato and PGP to get statistically significant results. 

In Figs.~\ref{fig:ranking_advg} and~\ref{fig:ranking_pgp}, AT gives more accurate ranking results, compared to other algorithms. 
In Advogato, the Kendall's tau correlation coefficients of AT are always greater than $0$.
Nearly 20\% of the ranking results are exactly the same (with a coefficient of $1$) as the ground truth.
In PGP, AT generates $>0.1$ Kendall's tau ranking correlation coefficients, and about 40\% of the ranking results are the same as the ground truth.
%
%
On the other hand, for ET and TR algorithms, only 20\% (Advogato) and 10\% (PGP) of their rankings are moderately correct, with coefficients $> 0.5$.
In other words, ET and TR do not work well in ranking users in an OSN, based on their trustworthiness.

\section{Related Work} 
\label{CH:RW}

\section{Definitions of Trust}
Trust has been widely studied in psychology, sociology and management domains. 
A widely accepted definition of trust was summarized by Rousseau in~\cite{rousseau1998not}, based on a cross-disciplinary literature review:
``Trust is a psychological state comprising the intention to accept vulnerability based upon positive expectations of the intentions or behaviors of another.''
Despite the various definitions of trust~\cite{Gefen:2003:TTO:2017181.2017185, Gambetta1988-GAMTMA, mcknight2002developing}, they are similar to Rousseau's, \ie, it can be concluded that trust is composed of two parts: expectation and vulnerability. 
While the former indicates the probability that the trustee will behave as expected, the latter shows the trustor's willingness of relying on the trustee. 
Specifically, the word vulnerability emphasizes the trustor's concerns about the uncertainty~\cite{doney1997examination, moorman1992relationships} of the trustee's future behaviors.
The definition of trust in this dissertation is inspired by the above studies, and we define trust as the probability that the trustee will behave as expected, from the perspective of the trustor.

Although trust is commonly confused with reputation, they are two different concepts.
Previous works~\cite{ganesan1994determinants, doney1997examination, jarvenpaa1999consumer} have identified the positive correlations between reputation and trust.
However, reputation is not equivalent to trust. 
According to the definition from Merriam-Webster dictionary and Wikipedia, reputation is the common opinion that people have about someone or something, \ie, the overall quality or character as seen or judged by people in general. 
In essence, reputation comes from the public and general opinion.
However, trust comes from individual opinions, \ie, from a trustor to a trustee with emphasis on personal interactions. 
On the other hand, reputation is a summary of past events while trust is the intention and expectation of the future.   

How to model the trust between users in OSNs has attracted much attention in recent years.
Existing trust models can be categorized into four groups: topology based, PageRank based, probability based, and subjective logic based models.  
In this section, we briefly introduce these works.

\subsection{Topology based Models}
Trust between users in OSNs was first studied by analyzing the characteristics of the network topology between the users. 
Topology based trust models treat a social network as a graph, where the edge between users represents the trust relation between them.
The advantage of these models is that they leverage random walk to evaluate users' trust, and thus can easily be applied in large-scale social networks.
By analyzing network topologies, the works in~\cite{wei2012sybildefender, danezis2009sybilinfer, yu2008sybilguard, 5313843} are able to identify untrustworthy nodes in an OSN.
The basic idea is to distinguish untrustworthy regions from trustworthy regions in a social network.
Specifically, the random walk algorithm is applied, starting from a trustor and searching the network to compute the probability that a trustee will be visited.
A low probability indicates the trustee is in the untrustworthy region, and thus untrustworthy.
Later on, the trust relation between two users is treated as a probabilistic value in~\cite{dubois2009rigorous}.
Then, indirect trust inference becomes a network reachability problem.
For example, in~\cite{zuo2009trust}, a social network is considered a resistor network where the resistance of each edge is derived from the trust of the two users being connected by the edge. 
In~\cite{golbeck2005computing,zhang2006social}, a depth first search algorithm is employed to compute the trust between two users.

\subsection{PageRank based Models}
Instead of analyzing the entire structure of a social network, PageRank based solutions are inspired by the assumption that trustworthy users are likely to have more connections from other users.
PageRank based trust models employ the PageRank algorithm~\cite{page1999pagerank} to compute the relative trust of users. 
For example, the EigenTrust algorithm\cite{Kamvar:2003:EAR:775152.775242} searches a social network based on the following rule: it moves from a user to another with probability proportional to how the user trusts the other, \ie, higher the trust, higher the probability.
In this way, EigenTrust algorithm is more likely to reach more trustworthy users.
Similarly, the TrustRank algorithm~\cite{Gyongyi:2004:CWS:1316689.1316740} also employs the PageRank algorithm to rank users, based on their relative trust values.
Both EigenTrust and TrustRank extend the PageRank algorithm that was commonly used to determine the importance scores of web pages.
The assumption that PageRank algorithm depends on, however, may not be realistic in a social network environment, leading to inaccurate trust assessment results.

\subsection{Probability based Models}
Unlike the aforementioned models that treat trust as either binary or real numbers, the probability based model considers trust as a probability, \ie, the likelihood that a trustee is trustworthy. 
Probability based trust models usually represent trust as a probability distribution.
In these models, a trustor uses its interactions with a trustee to construct a probabilistic distribution to estimate the trustee's trust. 
The benefit of these models is that trust is accurately modeled by a rich set of statistical and probability techniques, including Hidden Markov chain and maximum likelihood estimation.
In this category, trust can be represented by several different probability distributions~\cite{despotovic2005probabilistic, teacy2012efficient,elsalamouny2010hmm,liu2012modeling,vogiatzis2010probabilistic}. 
The trust assessment becomes the problem of likelihood estimation, regarding to the corresponding distribution's parameters, based on observed evidence.
For example, trust was first modeled as a binomial distribution in~\cite{josang2001logic}, and the likelihood estimation was carried out, based on Beta distribution.
Then, trust is considered a continuous random variable~\cite{despotovic2005probabilistic, teacy2012efficient}, and Gaussian distribution is used to model trust.
Binomial distribution is further extended to a multinomial distribution, to handle the cases where trust is a discrete random variable~\cite{fung2011dirichlet}.
Based on multinomial distribution, Bayesian inference~\cite{teacy2012efficient,despotovic2005probabilistic} and Hidden Markov Model (HMM)~\cite{elsalamouny2010hmm,liu2012modeling,vogiatzis2010probabilistic} can be applied to realize trust assessments.

While the former integrates evidence from various sources, \eg, reputation and preference similarity of users, the latter focuses on trust modeling in a dynamic environment.

\subsection{Subjective Logic based Models}
The 3VSL model extends the subjective logic (SL) trust models~\cite{josang2001logic, josang2005semantic,josang2008optimal} by redefining the uncertainty in trust, and thus achieves more accurate trust assessments.
Therefore, it is worth studying the fundamental principles of SL, and the limitations when SL is applied in trust assessment in OSNs.
Considering trust as a binary event, \ie, a trustee is trustworthy or untrustworthy, SL assumes the probability of a trustee being trustworthy follows Beta distribution.
The Beta distribution can be formed based on the amounts of positive and negative evidence, collected by a trustor about the trustee.
The advantage of SL based models is that trust is more accurately modeled and the uncertainty in trust is considered.
%
In~\cite{wang2006trust,wang2011probabilistic,wang2010evidence,kuter2010using, hang2009operators, hang2010trust}, the SL model is further refined and better trust assessment performance is achieved.

%
Subjective logic based models treat trust as an opinion and introduce a set of opinion operations, \eg, discounting and consensus operations to account for trust propagation and trust fusion, respectively. 
The consensus operation provides a mechanism to combine possibly-conflicting opinions to generate a consensus opinion~\cite{josang2002consensus}. 
%
On the other hand, the discounting operation is used to help a trustor to derive indirect trust of the trustee, based on other users' recommendations~\cite{josang2006exploring}.
For instance, if Alice trusts Bob, and Bob trusts Claire, then Alice will have an indirect opinion on Claire's trust. 
With the discounting and consensus operations, the trust between any two uses in an OSN can be computed.
In addition to the basic discounting and consensus operations, multiplication, co-multiplication, division, and co-division of opinions are also defined in the SL models~\cite{josang2005multiplication}.

%
Later on, the SL model is extended to support conditional inference~\cite{josang2008conditional}.
A conditional inference is represented in the form of ``IF $x$ THEN $y$'' where $x$ denotes the antecedent and $y$ the consequent proposition.
The antecedent $x$ is modeled by the SL model, so it is not a binary value; instead, it is a vector, representing the probability that this antecedent is true.
Overall, SL was proven to be compatible with binary logic, probability calculus, and classical probabilistic logic~\cite{josang2007probabilistic}.
The properties are also inherited in the proposed 3VSL model, which makes 3VSL computationally effective for trust assessments in OSNs.

\subsection{Applications of Trust in Online Systems}
Along with the rapid development of the Internet and online services, trust has been used in many applications for either improving users' quality of experience (QoE) or preventing the disturbance of malicious users. 
In this section, we briefly introduce these applications. 

\subsection{Trust in Cloud Computing}
Recently, trust was introduced in the concept of social cloud~\cite{8258338, 6682915, zhou2017will, 8622444, 8621918, assefi2016measuring, moyano2013re, assefi2015experimental, DiPietro201528}. 
In~\cite{6682915}, Mohaisen \textit{et al.} employ trust as a metric to identify good workers for an outsourcer through her social network. 
In~\cite{moyano2013re}, Moyano \textit{et al.} proposed a framework to employ trust  and reputation for cloud provider selection. 
In~\cite{DiPietro201528}, Pietro \textit{et al.} proposed a multi-round approach, called AntiCheetah, to dynamically assign tasks to cloud nodes, accounting for their trustworthiness. 
In~\cite{zhou2017will}, Zhou \textit{et al.} studied the trust factors affecting potential partnerships in bike sharing systems. 

\subsection{Trust in P2P Network and Semantic Web}
Trust analysis was first implemented in peer-to-peer (P2P) networks~\cite{4118688, Kamvar:2003:EAR:775152.775242, 1318566, 7346861}. In P2P networks, trust is used to evaluate the trustworthiness of a particular resource owner, and thereby identify malicious sources. Trust analysis was also applied to semantic webs~\cite{semantic1, semantic2, Artz200758}. The purpose of analyzing trust in semantic webs is to study the trustworthiness of data with efficient knowledge processing mechanisms. For example, the trustworthiness of web hyperlinks are studied in~\cite{1517920, Gyongyi:2004:CWS:1316689.1316740, Kleinberg:1999:ASH:324133.324140}. Trust analysis is then applied to filter untrustworthy contents in~\cite{Ding05onhomeland, Ding:2003:TBK:946251.947027, Bizer05o.:the, Downey:2005:PMR:1642293.1642459, Clarke:2001:ERQ:383952.384024, golbeck2009trust}. Finally, trust was used to evaluate the quality of contents on semantic webs in~\cite{Zhu:2000:IQM:345508.345602, Riloff:2005:ESC:1619499.1619511, Stoyanov:2005:MQA:1220575.1220691, Gil:2002:TIS:646996.711278,Golbeck:2005:CAT:1104446, 1517920}. 

\subsection{Trust in Cyber-Physical and Edge Computing Systems}
Trust analysis is also introduced in cyber-physical systems (CPS), \eg, wireless sensor networks~\cite{6180964, 6142264, 8302985, 6040313, liu2019novel, Chen:2017:HEW:3136518.3047646, 7786108, 7236522, 7444932, chen2019understanding} and vehicular networks~\cite{SAT, 8567683, chen2019cooper, chen2019f}. For example, a trust based framework is proposed to secure data aggregation in wireless sensor networks~\cite{sensor}, which evaluates the trustworthiness of each sensor node by the Kullback-Leibler~(KL) distance to identify the compromised nodes through an unsupervised learning technique. In~\cite{4481349}, trust analysis is employed to identify malicious and selfish nodes in a mobile ad hoc network. In addition, Xiaoyan \textit{et al}. propose a new trust architecture, called situation-aware trust (SAT), to address several important trust issues in vehicular networks, which are essential to overcome the weaknesses of current vehicular network security and trust models~\cite{SAT}.

\subsection{Trust in Spam Detection and Sybil Defense}
Another important domain in which trust analysis is widely applied is Sybil defense and  spam detection~\cite{6547122, wei2012sybildefender, Gao:2014:SAD:2664243.2664251, Hu:2013:SSD:2540128.2540508, Tan:2013:UUS:2505515.2505581, 5934998}. The goal of these works is to identify forged multiple identities and spam information in OSNs. 
The basic idea of~\cite{6547122, wei2012sybildefender} is to employ random walk to rank the neighbors in a given OSN from a seed node, and extract a trust community composed of high ranking nodes.  
Then, the users outside the trust community will be considered as not trustworthy, \ie, potential Sybil nodes. %
In~\cite{Tan:2013:UUS:2505515.2505581}, Tan \textit{et al.} integrated traditional Sybil defense techniques with the analysis of user-link graphs. 
In~\cite{5934998}, Mohaisen \textit{et al.} proposed a derivation of the random walk algorithm, which employs biased random mechanism, to account for trust and other social ties. In~\cite{6553246}, besides graph based features, Yang \textit{et al.} introduced some other features to identify spammers.  
In addition, in~\cite{Hu:2013:SSD:2540128.2540508, Gao:2014:SAD:2664243.2664251}, spam detection approaches based on user similarity and content analysis are studied. 

\subsection{Trust in Recommendation and Crowdsourcing Systems}
In addition to Sybil defense in OSNs, trust analysis is also useful in recommendation systems~\cite{Basu:2014:OPF:2670967.2670968, Zou:2013:BPA:2505515.2507875, Andersen:2008:TRS:1367497.1367525,hang2010trust,Jamali:2010:MFT:1864708.1864736, massa2007trust}.
For example, in~\cite{Zou:2013:BPA:2505515.2507875}, Zou \textit{et al.} proposed a belief propagation algorithm to identify untrustworthy recommendations generated by spam users.
In~\cite{Basu:2014:OPF:2670967.2670968}, Basu \textit{et al.} proposed a privacy preserving trusted social feedback scheme to help users obtain opinions from friends and experts whom they trust. 
In~\cite{Andersen:2008:TRS:1367497.1367525}, Andersen \textit{et al.} proposed a trust-based recommendation system that generates personalized recommendations by aggregating the opinions from other users. 
In addition, five axioms about trust in a recommendation system are studied in~\cite{Andersen:2008:TRS:1367497.1367525}.

\section{Conclusions}
\label{conclusion}

In this paper, the three-valued subjective logic is proposed to model and compute trust between any two users connected within OSNs. 
3VSL introduces the uncertainty space to store evidence distorted from certain spaces as trust propagates through a social network, and keeps track of evidence as multiple trusts combine. 
We discover that there are differences between distorting and original opinions, \ie, distorting opinions are so unique that they can be reused in trust computation while original opinions are not.
This property enables 3VSL to handle complex topologies, which is not feasible in the subjective logic model.

Based on 3VSL, we design the AT algorithm to compute the trust between any pair of users in a given OSN. 
By recursively decomposing an arbitrary topology into a parsing tree, we prove AT is able to compute the tree and get the correct results.    
An open issue to 3VSL and OpinionWalk is how to estimate the value of evidences.
This issue is further studied and addressed by probabilistic graphic models or neural network models~\cite{8737469}.

We validate 3VSL both in experimental evaluations. 
The evaluation results indicate that 3VSL is accurate in modeling computing trust within complex OSNs. 
We further compare the AT algorithm to other benchmark trust assessment algorithms.
Experiments in two real-world OSNs show that AT is a better algorithm in both absolute trust computation and relative trust ranking. 
%

\bibliographystyle{unsrt}
\bibliography{./bib/output}

\end{document}